\documentclass[
aps,
prc,
notitlepage,
citeautoscript, superscriptaddress,
reprint,
eqsecnum
]{revtex4-2}

\usepackage[dvipsnames,usenames]{xcolor}
\usepackage[caption=false]{subfig}
\usepackage{graphicx}
\usepackage{dcolumn}
\usepackage{bm}
\usepackage{mathrsfs,natbib}
\usepackage[title]{appendix}
\usepackage{graphicx,epsf,amssymb,amsbsy,amsfonts,amssymb,amsmath}
\usepackage{slashed}
\usepackage{comment}
\usepackage{todonotes}

\usepackage{color,soul}

\newcommand{\be}{\begin{equation}}
\newcommand{\ee}{\end{equation}}
\newcommand\beq{\begin{eqnarray}}
\newcommand\eeq{\end{eqnarray}}

\newcommand\ket[1]{| #1 \rangle}

\newcommand{\mybar}[1]

\usepackage[colorlinks=true]{hyperref}
\hypersetup{
    unicode=false,          
    pdftoolbar=true,        
    pdfmenubar=true,        
    pdffitwindow=false,     
    pdfstartview={FitH},    
    pdftitle={},    
    pdfauthor={},     
    pdfsubject={},   
    pdfcreator={},   
    pdfproducer={}, 
    pdfkeywords={} {} {}, 
    pdfnewwindow=true,      
    colorlinks=true,       
    linkcolor=blue, 
    citecolor=blue,        
    filecolor=blue,      
    urlcolor=blue           
}

\begin{document}

\title{Floquet insulators and lattice fermions beyond naive time discretization}
\author{Thomas Iadecola}
\email{iadecola@iastate.edu }
\affiliation{
Department of Physics and Astronomy, Iowa State University, Ames, Iowa 50011, USA
}%
\affiliation{Ames National Laboratory, Ames, Iowa 50011, USA}%
\author{Srimoyee Sen}%
\email{srimoyee08@gmail.com}
\affiliation{
Department of Physics and Astronomy, Iowa State University, Ames, Iowa 50011, USA
}%
\author{Lars Sivertsen}%
\email{lars@iastate.edu }
\affiliation{
Department of Physics and Astronomy, Iowa State University, Ames, Iowa 50011, USA
}%

\date{\today}

\begin{abstract}
Periodically driven quantum systems known as Floquet insulators can host topologically protected bound states known as ``$\pi$ modes" that exhibit response at half the frequency of the drive.
Such states can also appear in undriven lattice field theories when time is discretized as a result of fermion doubling, raising the question of whether these two phenomena could be connected.
Recently, we demonstrated such a connection at the level of an explicit mapping between the spectra of a continuous-time Floquet model and a discrete-time undriven lattice fermion model.
However, this mapping relied on a symmetry of the single-particle spectrum that is not present for generic drive parameters.
Inspired by the example of the temporal Wilson term in lattice field theory, in this paper we extend this mapping to the full drive parameter space by allowing the parameters of the discrete-time model to be frequency-dependent. 
The spectra of the resulting lattice fermion models exactly match the quasienergy spectrum of the Floquet model in the thermodynamic limit. Our results demonstrate that spectral features characteristic of beyond-equilibrium physics in Floquet systems can be replicated in static systems with appropriate time discretization.
\end{abstract}

\maketitle
\section{Introduction}
Periodically driven quantum systems can exhibit exotic behaviors not seen at equilibrium~\cite{BukovReview,CayssolReview,RudnerReview,ElseReview,SachaReview,KhemaniReview}. Even though periodic driving in these systems leads to non-conservation of energy, one can still define an analogous quantity, the quasienergy, related to the eigenvalue spectrum of the evolution operator over a driving period $T$. Quasienergy is conserved modulo $2\pi/T$ and can therefore be viewed as a periodic variable akin to crystal momentum, leading to distinctive spectral properties absent in undriven Hamiltonians. 
As an example, consider equilibrium fermion topological insulators (TIs) and superconductors (TSCs)~\cite{HasanReview}.  Localized zero-energy boundary modes known as zero modes appear when such systems are defined in finite volume with open boundary conditions (OBC).
In two or more spatial dimensions, they are associated with massless modes that propagate on the boundary, while in one spatial dimension they are fully localized at the two ends of the system~\cite{TeoKane}. Periodically driven analogs of equilibrium TIs known as Floquet insulators can also have boundary modes when placed in finite volume with OBC~\cite{CayssolReview,RudnerReview}. However, these boundary modes can be zero modes or so called $\pi$ modes, having quasienergy $0$ or $\pi/T$, respectively~\cite{Thakurathi,Rudner13,Khemani,ElseTC,ElseFSPT,vonKeyserlingk}. The appearance of $\pi$ modes is a uniquely non-equilibrium phenomenon that cannot be seen in equilibrium quantum systems with continuous time. 

However, something analogous may happen even for a static system when time is discretized. 
This is well known in fermion lattice field theories and goes by the name of fermion doubling \cite{Nielsen:1980rz, Nielsen:1981xu, GuyMoore,Ambjorn:1990pu,Aarts:1998td, Mou:2013kca}. 
In lattice field theory typically space and time are both discretized. 
A naive discretization of spacetime in fermionic theories leads to a doubling of fermion species---i.e., for $d$ discretized spacetime dimensions, the fermion operator experiences a $2^d$ fold increase in the number of degenerate eigenstates. 
These degenerate eigenstates are ``$\pi$-paired" with each other.
In other words, for a cubic spacetime lattice with spacing $a$, if the state with $d$-momentum $(p_0, p_1, \cdots, p_{d-1})$ is an eigenstate of the fermion operator with eigenvalue $f_p$, then so are the states $(\pi/a-p_0,  p_1, \cdots, p_{d-1})$, $(p_0, \pi/a- p_1, \cdots, p_{d-1})$, $\cdots$, $(\pi/a-p_0,\pi/a-p_1\cdots,\pi/a-p_{d-1})$. 
This has an interesting implication for a continuous-time theory with boundary zero modes. 
When such a theory is considered on a discrete time lattice with lattice spacing $\tau$, one automatically gains an extra mode, i.e. the $\pi$ mode with energy $\pi/\tau$ localized on the boundary. 

The presence of $\pi$ modes in lattice field theory naturally begets comparison with Floquet insulators.
In fact, one can ask if the non-equilibrium Floquet insulator spectrum can be 
replicated with a discrete-time theory 
for a time-independent Hamiltonian. In Ref.~\cite{FloquetLattice1}, we showed that this is indeed possible for a particular $1+1$ dimensional Floquet insulator model restricted to a certain line in the space of drive parameters. Our demonstration relied on the observation that, along this line, the quasienergy spectrum is fully $\pi$-paired: for every quasienergy value $\epsilon_i$, there is a corresponding one at $\pi/T - \epsilon_i$. This motivated us to map the Floquet spectrum to that of a static fermion Hamiltonian with naive time discretization of lattice spacing $T$. We found that the corresponding static Hamiltonian can be that of a Wilson-Dirac theory or of a Su-Schrieffer-Heeger (SSH) model~\cite{SSH}.

However, there are other parts of this model's phase diagram that do not exhibit $\pi$ pairing. In these regions, with OBC, the boundary modes at quasienergies $0$ and $\pi/T$ need not always appear together as they do for naively discretized lattice fermions. These regions of the phase diagram therefore have no analogs in naively discretized fermion theories. 
However, there exist several schemes for removing fermion doubling on the lattice. 
In this paper we focus on Euclidean lattice field theories.
There, one can introduce a Wilson term in the fermion Lagrangian to remove fermion doubling.
For other methods see~\cite{Kaplan:1992bt, Shamir:1993zy, PhysRevD.25.2649, Neuberger:1997fp, Neuberger:1998wv, PhysRevD.11.395}. 
In a lattice fermion theory exhibiting both a zero-frequency and a $\pi/T$-frequency mode, a Wilson term can be engineered to remove either of the two. 
In standard lattice field theory, it is typically the $\pi$ mode that is removed. 
The Wilson fermion approach relies on introducing a frequency dependence in the fermion mass term, which can also be thought of as adding a higher dimensional operator involving time derivatives in the lattice fermion action.
This frequency-dependent mass breaks the degeneracy of eigenvalues for the fermion operator between the zero and $\pi$ modes. 
Inspired by this mechanism, in this paper we map the entire phase diagram of the Floquet insulator model studied in Ref.~\cite{FloquetLattice1} to that of a discrete-time lattice fermion theory with a time-independent Hamiltonian. The action of our target theory written in frequency space includes frequency-dependent parameters. In particular it includes a frequency-dependent mass term, just as in the case of the standard Wilson term. We construct two such examples: we call the first a modified Wilson-Dirac theory and the second a modified SSH model.  

The organization of the paper is as follows. We begin with background discussion in Sec.~\ref{sec: Background}. In Sec.~\ref{sec: Model} we briefly describe the Floquet model in question and its spectrum. In Sec.~\ref{sec: Lattice} we review how Wilson terms can be used to circumvent fermion doubling in lattice fermion theories. In Sec.~\ref{sec: Floquet-to-lattice} we formulate the Euclidean spacetime version of the mapping between quasienergy and discrete-time spectra found in Ref.~\cite{FloquetLattice1} for the $\pi$-paired region of the Floquet model's phase diagram. In Sec.~\ref{sec: Off-axis} we discuss how to formulate such a mapping throughout the entire phase diagram irrespective of $\pi$ pairing. We will first explain our construction in frequency space, and then take the Fourier transform to discuss the structure of the lattice theory in the time domain. 
In Sec.~\ref{sec: Conclusion} we conclude and provide an outlook for future research.

\section{Background}
\label{sec: Background}
\subsection{Model and Phase Diagram}
\label{sec: Model}
We begin with the SSH model in $1+1$ dimensions 
on a spatial lattice of $2N$ sites.
The SSH Hamiltonian has the form 
\beq
H_{\text{SSH}}=\frac{v}{2} H_0+\frac{u}{2} H_1,
\label{ssh}
\eeq
where 
\begin{align}
\begin{split}
H_0&=2\sum^{N-1}_{j=0}(a_{2j}^{\dagger}a_{2j+1}+\text{H.c.}),\\
H_1&=2\sum^{N-1}_{j=0}
(a_{2j+1}^{\dagger}a_{2j+2}+\text{H.c.}),
\label{h}
\end{split}
\end{align}
where $a_i$ is the fermion annihilation operator on site $i=0,\dots,2N-1$. The parameters $u,v$ are both assumed to be positive, without loss of generality. Unless otherwise specified, we consider periodic boundary conditions (PBC) such that $a_{2N}\equiv a_0$. The energy spectrum of this model with PBC is given by \begin{align}
\label{essh}
    E_{\mathrm{SSH},\pm}(p_1) = \pm\sqrt{u^2+v^2+2uv\cos(2p_1)},
\end{align}
where $0\leq p_1 < \pi$ is the crystal momentum and we have set the spatial lattice spacing to 1. Expanding this expression about $p_1=\pi/2$ yields the dispersion of a Dirac fermion with mass $\propto u-v$. The spectrum with PBC is invariant under exchange of $u$ and $v$ as it depends only on the square of the mass. However, the spectrum with OBC is not. This is because the two different choices for $u$ and $v$ correspond to two different equilibrium topological phases, one of them trivial and the other nontrivial~\cite{SSH,JackiwRebbi}.
The trivial and nonstrivial phases are characterized by the absence or presence, respectively, of zero-energy modes localized at the edges of the chain. With the SSH model defined as in Eqs.~\eqref{ssh} and \eqref{h}, the topological phase is the one for which $u>v$.

\begin{figure}[t!]
\includegraphics[width = 0.7\columnwidth]{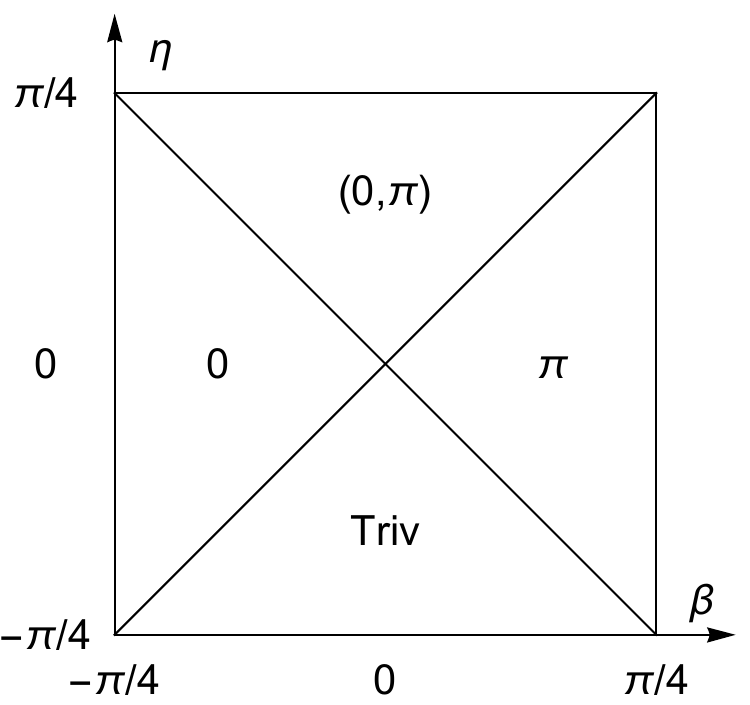}
\caption{Illustration of the Floquet phase diagram as a function of the drive parameters $\beta$ and $\eta$ defined in Eq.~\eqref{beta-eta}. The different phases are distinguished by whether they exhibit boundary zero or $\pi$ modes, or both $(0,\pi)$, or neither (Triv). The diagonal lines $|\beta|=|\eta|$ indicate gap closings in the single-particle quasienergy spectrum.}\label{phase_diagram}
\end{figure}

The static SSH model \eqref{ssh} allows us to define the Floquet model discussed in Ref.~\cite{FloquetLattice1}. The model is a complex-fermion version of the Floquet-Majorana model (or its dual, the kicked Ising model) studied in Ref.~\cite{vonKeyserlingk,Khemani,Potter}. It is defined by the time-evolution operator\begin{align}
U(t)&=\begin{cases}e^{-i H_0 t} & \text{for } 0<t<t_0\\
e^{-i H_1(t-t_0)}e^{-i H_0 t_0} & \text{for } t_0\leq t<t_0+t_1.
\end{cases},
\label{u}
\end{align}
In order to extract the quasienergy spectrum we have to consider this evolution operator at integer multiples of the drive period $T=t_0+t_1$. This allows us to define the Floquet Hamiltonian given by 
\beq
\label{eq:HF}
H_F=\frac{i}{T}\ln[U(T)].
\eeq
The quasienergies are the eigenvalues of $H_F$ and are conserved up to $2\pi/T$. Defining the variables
\begin{equation}
\label{beta-eta}
    \beta = \frac{t_0}{T}-\frac{\pi}{4},\quad\quad \eta = \frac{t_1}{T}-\frac{\pi}{4},
\end{equation}
we illustrate the Floquet phase diagram as a function of $\beta$ and $\eta$ in Fig.~\ref{phase_diagram}. With PBC, the quasienergy eigenvalues can be found by transforming $U(T)$ to momentum space. They are given by
\begin{align}
\epsilon_{\pm}&(p_1) 
= \pm\frac{1}{T}\arccos\Big\{\frac{1}{4}\big[-\cos(2 p_1 -2\beta-2\eta)
\nonumber
\\
&+2 \cos(2\beta-2\eta)-\cos(2p_1+2\beta-2\eta)
\nonumber
\\
&-\cos(2p_1 - 2\beta +2 \eta)-2\cos(2\beta+2\eta)
\nonumber
\\
&\quad\quad\quad\quad\quad\quad-\cos(2p_1 +2\beta+2\eta)\big]\Big\}.\label{theta}
\end{align}
The eigenvalues are invariant under an exchanging $\beta$ and $\eta$ and exhibit periodicity such that we can restrict $-\pi/4<\beta\leq \pi/4, -\pi/4\leq\eta\leq\pi/4$ to study the full phase diagram. We observe that the quasienergy gap is nonzero everywhere except at $\eta=|\beta|$. These gapless lines divide the region $-\pi/4<\beta<\pi/4$, $-\pi/4<\eta<\pi/4$ into four different phases. These four phases are called the trivial phase, the $0$ phase, the $\pi$ phase and the $(0,\pi)$ phase. The names correspond to the boundary spectrum in these regions with OBC. The trivial phase does not have have any boundary modes, the zero phase has zero modes, the $\pi$ phase has $\pi$ modes, and the $(0,\pi)$ phase has zero and $\pi$ modes localized at the edges. In the next section we describe how some of these same features can arise in lattice fermion theories in discrete spacetime. Since lattice field theory is typically formulated in Euclidean space, we will do so in our treatment.

\subsection{Wilson Terms in Lattice Fermion Theories}
\label{sec: Lattice}
We begin by reviewing fermion doubling in Euclidean lattice field theory before discussing the connection to Floquet systems. We consider the following Euclidean spacetime Dirac fermion Lagrangian on a $1+1$ dimensional spacetime lattice:
\beq
\mathcal{L}=\bar{\psi}(\gamma_{\mu}\nabla_{\mu}+m)\psi,
\eeq
where $\mu=0,1$ with $\mu=0$ being the Euclidean time direction, $\nabla_{\mu}$ is the symmetric discrete derivative given by 
$(\nabla_{\mu})_{x,x'}=(\delta_{x',x+a_\mu}-\delta_{x',x-a_\mu})/(2a_{\mu})$, and $a_{\mu}$ is the lattice spacing in the $\mu$ direction. We have also adopted the Einstein summation convention over repeated Greek indices. $\gamma_{\mu}$ are Euclidean gamma matrices satisfying $\{\gamma_{\mu},\gamma_{\nu}\}=2\delta_{\mu\nu}$.
In Fourier space the Lagrangian is of the form 
\beq
\mathcal{L}=\bar{\psi}[-i\gamma_{\mu}(a_{\mu})^{-1}\sin (p_{\mu}a_{\mu})+m]\psi.
\eeq
The object in square brackets above is sometimes referred to as the fermion or Dirac operator.
When the mass $m=0$, the spectrum of the Dirac operator has four zeroes at 
$(p_0,p_1)=(0,0)$, $(0,\pi/a_1)$, $(\pi/a_0,0)$, and $(\pi/a_0,\pi/a_1)$. The zeroes other than the one at $(0,0)$ are called doublers since their appearance is a consequence of fermion doubling.

For $m\neq 0$, the states corresponding to these zeroes become degenerate eigenstates of the fermion operator with the same eigenvalue $m$. 
This is undesirable in lattice simulations and there are several techniques to remove the extra modes.
One of the standard techniques to remove the doublers is to add a Wilson term to the Lagrangian which shifts the mass term from $m$ to $m-\frac{R_{\mu} (a_\mu)^2}{2}\nabla^2_\mu$
where $\nabla^2_{\mu}=(\delta_{x',x+a_\mu}+\delta_{x',x-a_\mu}-2\delta_{x',x})/(a_{\mu}^2)$ is the discrete second derivative and $R_{\mu}$ is the Wilson parameter. The corresponding Lagrangian is known as the Wilson-Dirac Lagrangian and here the eigenvalue degeneracy of the four states at $(p_0,p_1)=(0,0)$, $(0,\pi/a_1)$, $(\pi/a_0,0)$, and $(\pi/a_0,\pi/a_1)$
is lifted. \footnote{In theories with Euclidean rotational invariance, $R_{\mu}$ is $\mu$ independent and set to $R_{\mu}=R$.}

The full Wilson-Dirac action therefore is of the form 
\beq
\mathcal{S}&=\int d^2x \,\,\bar{\psi}\,G_{\text{WD}}\,\psi,
\eeq
where $G_{\text{WD}}$ is the Wilson Dirac operator,
\beq
\label{gwd}
G_{\text{WD}}&=&\gamma_{\mu}\nabla_{\mu}+m-\frac{R_{\mu}a_{\mu}^2}{2}\nabla_{\mu}^2,\nonumber\\
&&\,\,\,\,\,\,\,\text{or}\\
&=&-i\gamma_{\mu}(a_{\mu})^{-1}\sin(p_{\mu}a_\mu)+m+R_{\mu}[1-\cos(p_{\mu}a_{\mu})],\nonumber
\eeq
in position and momentum space (with PBC), respectively.
Let us now consider the eigenvalues of the Wilson-Dirac operator, first with PBC. For an eigenstate $|p_0,p_1\rangle$ of $G_{\rm WD}$ with frequency $p_0$ and momentum $p_1$, the eigenvalue is 
\begin{widetext}
\beq
g_{\pm}(p_0,p_1)=m+R_0\left[1-\cos (a_0p_0)\right]+R_1\left[1-\cos (a_1p_1)\right]\pm i\sqrt{\frac{1}{a_0^2}\sin^2 (a_0 p_0) + \frac{1}{a_1^2}\sin^2  (a_1p_1)}.
\eeq
\end{widetext}
Clearly, the degeneracy of the states $|p_0, p_1\rangle$, $|\pi/a_0 - p_0, p_1\rangle$, $|p_0, \pi/a_1 - p_1\rangle$ and $|\pi/a_0 - p_0, \pi/a_1 - p_1\rangle$ is broken (even though there may be accidental degeneracies for some parameter values).

Let us now understand how boundary zero and $\pi$ modes arise in this theory and how their degeneracy is affected by the Wilson term. To do this, we must consider the spectrum of the Wilson-Dirac operator with OBC in space. Our goal is to identify localized eigenstates of the Wilson-Dirac operator with zero eigenvalue, of frequency $p_0=0$ (zero mode) and $\pi/a_0$ ($\pi$ mode). To search for these modes, we can set $\gamma_0\nabla_0\rightarrow 0$ inside $G_{\text{WD}}$ and solve for the zero eigenvalues of the rest of the operator. 
Without loss of generality, we first pick $m=-1/a_1, R_1=1/a_1$. If we now set $R_0=0$, we find four spatially localized zero-eigenvalue eigenstates of $G_{\text{WD}}$, two of them with frequency $p_0=0$ (one on each edge) and two others with $p_0=\pi/a_0$ (one on each edge), indicating $\pi$ pairing. If now we set $R_0=1/a_1$, the system acquires an effective frequency-dependent mass $m_{\rm eff}(p_0a_0)=m+R_0[1-\cos(p_0a_0)]=-\frac{1}{a_1}\cos(p_0 a_0)$. Since the Wilson-Dirac operator only has localized zero-eigenvalue boundary modes when the $m_{\rm eff}/R_1<0$, we now find that the existence of such a mode depends on $p_0$. In particular, when $p_0=0$, $m_{\rm eff}(0)=-1/a_1$ and a localized zero-eigenvalue solution exists, whereas when $p_0=\pi/a_0$, $m_{\rm eff}(\pi)=1/a_1$ and no zero-eigenvalue solution exists. Next, we can consider $m=1/a_1$, $R_1=1/a_1$, and $R_0=-1/a_1$. Now, the effective frequency dependent mass $m_{\rm eff}(p_0a_0)=\frac{1}{a_1}\cos(p_0 a_0)$, so the situation is reversed: we obtain zero-eigenvalue boundary solutions for $p_0=\pi/a_0$ but not $p_0=0$. Therefore, we see that the presence of boundary zero and $\pi$ modes can be tuned by introducing appropriate Wilson terms.

\subsection{Review of the Floquet-to-Lattice Mapping for $\beta=0$}
\label{sec: Floquet-to-lattice}
The above discussion shows that $\pi$ pairing in the lattice fermion spectrum is guaranteed for $R_0=0$ but absent for $R_0\neq 0$, which promotes the fermion mass to a frequency-dependent quantity. In Ref.~\cite{FloquetLattice1}, we showed how to match [exactly for PBC and up to $O(1/N)$ corrections for OBC] the quasienergy spectrum~\eqref{theta} onto the spectrum of a time-independent lattice fermion theory in the absence of any frequency dependence in the model parameters. In this case, the time direction is naively discretized such that only first-order time derivatives appear in the action and $\pi$ pairing (i.e., fermion doubling) is unavoidable. In the Floquet model, the quasienergy spectrum \eqref{eq:epsilon_beta0} is only $\pi$ paired along the line $\beta=0$ defined in Eq.~\eqref{beta-eta} that corresponds to a vertical cut at $t_0/T = \pi/4$ in the phase diagram in Fig.~\ref{phase_diagram}. Thus, our naive spectral mapping was restricted to this parameter line. In this section, we review this mapping before generalizing it in Sec.~\ref{sec: Off-axis} to the full Floquet phase diagram.

To define the mapping, we consider a Euclidean spacetime lattice with temporal lattice spacing $a_0=T$ (corresponding to the drive period of the Floquet model) and spatial lattice spacing $a_1=1$. For convenience, we repackage the fermion operator for a generic fermion theory on this lattice as $\gamma_0(\mathcal F+\mathcal{F}')$, where $\mathcal F\equiv\nabla_0$ and $\mathcal F'$ contains spatial derivatives and possibly higher-order time derivatives.
The corresponding Euclidean action is then
\beq
\mathcal{S}=\int d^2x\,\, \bar{\psi}\, \gamma_0(\mathcal F+\mathcal{F}')\, \psi.
\eeq
As an example, for the Wilson-Dirac theory discussed in the previous section
\beq
\mathcal{F}'=\gamma_0\left(\gamma_i\nabla_i+m-\frac{R_{\mu}a_{\mu}^2}{2}\nabla^2_\mu\right).\label{F'}
\eeq
More generally, $\mathcal{F}'$ can have other forms and we initially remain agnostic about its exact form.  

We can now formulate the mapping between the Floquet and lattice fermion spectra by demanding that solutions to the equation 
\begin{align}
\label{eq:ff't}
i\mathcal{F}\ket{\psi(t,x)}=\mathcal{F}'\ket{\psi(t,x)},
\end{align}
with PBC match one to one to the solutions of the Floquet Schr{\"o}dinger equation given by $i\partial_t\ket{\psi(t,x)}=H_F\ket{\psi(t,x)}$, again with PBC~\footnote{Note that, in previous work, we had mapped the Floquet spectrum to Minkowski space lattice Fermions using the same condition. In that case, the zeroes of the Minkowski Fermion operator match one to one with the zeroes of the Floquet Schr{\"o}dinger operator $i\partial_t-H_F$. The solutions to the equation \eqref{eq:ff't} in Euclidean space are not necessarily zeroes of the fermion operator unless $i\mathcal{F}=\mathcal{F}'=0$. This is the only distinction between the Euclidean and Minkowski constructions and it does not affect any of our analysis.}. More concretely, we demand that the frequency values $p_0$ that solve the frequency-space version of Eq.~\eqref{eq:ff't},
\begin{align}
\label{eq:ff'p0}
\frac{1}{T}\sin(p_0 T)\ket{\psi(p_0,p_1)}=\mathcal{F}'(p_0, p_1)\ket{\psi(p_0,p_1)}.
\end{align}
match one-to-one with the quasienergies $\epsilon$. Here $\ket{\psi(p_0, p_1)}$ is the Fourier transform of $\ket{\psi(t,x)}$. 

For every crystal momentum $p_1$, Eq.~\eqref{eq:ff'p0} is satisfied by two different frequencies $p_0>0$ (and also two different $p_0<0$) irrespective of the details of $\mathcal{F}'$. For an example, see Appendix \ref{app:eigen}. 
In the case where $\mathcal F'$ is frequency-independent, the two positive frequencies become $\pi$ paired, i.e.~they sum to $\pi/T$. Similarly the negative ones sum to $-\pi/T$.
To see this, note that there are two solutions to Eq.~\eqref{eq:ff'p0} for every eigenvalue $f'$ of $\mathcal F'$ (corresponding to crystal momentum $p_1$), namely
\begin{align}
\label{eq:pi-paired mapping}
p_0 T=\sin^{-1}(f'T) \indent \text{and}\indent p_0 T=\pi-\sin^{-1}(f'T).
\end{align}
The absence of frequency dependence in $\mathcal{F}'$ allows us to interpret it as a lattice fermion Hamiltonian $H$ with energies $E=f'$. In this case, Eq.~\eqref{eq:ff't} is a discrete time Schr\"odinger equation with Hamiltonian $H$. 
Thus we seek to identify the quasienergies with the doubled spectrum of $H$ when placed on a discrete-time lattice.

Note that there are $2N$ solutions to the Floquet Schr\"odinger equation since there are $2N$ lattice sites in the Floquet model. 
If the solutions of Eq.~\eqref{eq:ff'p0} are to match one-to-one those of the Floquet problem, and if we assume that the degrees of freedom in $\mathcal F'$ are spinless fermions as in the original Floquet problem, then $H$ must be defined on a spatial lattice of $N$ sites so that it has $N$ eigenvalues. Fermion doubling then accounts for the ``missing" half of the Floquet spectrum. If we instead assume that $\mathcal F'$ describes spinful fermions, then we must consider a discrete-time lattice theory with $N/2$ sites, one quarter of those in the original problem, in order to obtain the correct number of eigenvalues after fermion doubling.

We now describe the construction of the lattice Hamiltonian $H$.
We begin by writing down the PBC quasienergy eigenvalues \eqref{theta} on the $\beta=0$ line:
\begin{align}
\label{eq:epsilon_beta0}
\epsilon_\pm(p_1) = \pm\frac{1}{T}\cos^{-1}[-\cos(2\eta)\cos(2p_1)]. 
\end{align}
Due to $\pi$ pairing, the operator $\sin(H_F T)$ exhibits eigenvalue degeneracy for the states with momenta $p_1$ and $p_1+\pi/2$.
Thus, we can split the quasienergy spectrum $\epsilon_\pm$ into two branches, $\epsilon^1_\pm$ and $\epsilon^2_\pm$, corresponding to momenta $\frac{\pi}{4}\leq p_1\leq \frac{3\pi}{4}$ and $-\frac{\pi}{4}\leq p_1<\frac{\pi}{4}$, respectively,
such that $\sin(\epsilon^2_\pm T)=\sin(\epsilon^1_\pm T)$.
Thus, we can restrict to, e.g., branch $1$ and demand that $\sin(\epsilon^1_\pm T)$ matches onto the energy spectrum of some time-independent $H$.
In Ref.~\cite{FloquetLattice1}, we performed this matching for two choices of $H$: an SSH Hamiltonian of the form~\eqref{ssh} defined on $N$ lattice sites or a Wilson-Dirac Hamiltonian of the form
\begin{align}
\label{hwd}
\begin{split}
    H_{\rm WD}&=\!\!\sum^{N-1}_{x,x'=0}\!\bar\psi_x\bigg[R_1\, \gamma^1\, (-i\nabla_{x,x'})\\
    &\qquad\qquad\qquad-\frac{R_1}{2}\, \nabla^2_{x,x'}+m\, \delta_{x,x'}\bigg]\psi_{x'},
\end{split}
\end{align}
defined on $N/2$ lattice sites (where $N$ is even) as the underlying fermions are spinful.
Here $\gamma^{\mu}$ correspond to the Minkowski gamma matrices which are related to the Euclidean gamma matrix using, $\gamma^0=\gamma_0$
$\gamma^1=i\gamma_1$. For either choice of $H$ the spectral matching requires solving for appropriate values of the parameters in the respective models.

In the SSH case, we perform the matching by demanding
\begin{align}
\label{esshp}
    E_{\rm SSH,\pm}(k) = \frac{1}{T}\sin\left[T\epsilon^1_\pm\left(\frac{p}{2}+\frac{\pi}{4}\right)\right],
\end{align}
where $0\leq p< \pi$.
The corresponding SSH Hamiltonian is given by Eq.~\eqref{ssh} with parameters
\begin{align}
 u=\frac{1\pm\sin(2\eta)}{2T},\indent\text{and}\indent v = \frac{1}{T}-u.
\end{align}
(Note that we have assumed $u,v\geq0$ for simplicity).
The sign ambiguity in $u$ can be resolved by demanding that the spectra also match in the thermodynamic limit with OBC for the same choice of parameters. This is accomplished when~\cite{FloquetLattice1}
\begin{align}
\label{uv}
 u=\frac{1+\sin(2\eta)}{2T},\indent\text{and}\indent v = \frac{1}{T}-u.
\end{align}
For the choice above, the SSH Hamiltonian has a zero mode with OBC for $\eta>0$. As a result the discrete-time spectrum has both a zero mode and a $\pi$ mode. Similarly, for $\eta<0$, the SSH Hamiltonian does not have a zero mode with OBC. Therefore the corresponding discrete time theory has neither a zero mode nor a $\pi$ mode. This captures the trivial and $(0,\pi)$ phases in Fig.~\ref{phase_diagram} along the line $\beta=0$.

To match onto the spectrum of the Wilson-Dirac Hamiltonian~\eqref{hwd}, 
\begin{align}
E_{\rm WD,\pm}(p) = \pm\sqrt{R_1^2 \sin^2 p+[m+R_1(1-\cos p)]^2},
\end{align}
with $-\pi\leq p<\pi$, we demand that
\begin{align}
\label{ewdp}
    E_{\rm WD}(p)=\frac{1}{T}\sin\left[T\epsilon\left(\frac{p}{4}+\frac{\pi}{2}\right)\right].
\end{align}
Solving this condition for $m$ and $R_1$ 
gives
\begin{align}
m=\pm\frac{\sin(2\eta)}{T}\indent \text{and}\indent R_1=\frac{1}{2T}-\frac{m}{2},
\end{align}
where we have assumed $R_1\geq0$ for simplicity. Again, we resolve the sign ambiguity in $m$ by demanding that the spectra also match in the thermodynamic limit for OBC. This gives~\cite{FloquetLattice1}
\beq
m=-\frac{\sin(2\eta)}{T},\,\, R_1=\frac{1}{2T}-\frac{m}{2},
\label{m}
\eeq
for which boundary zero modes appear when $\eta >0 $ and not for $\eta < 0$. The corresponding discrete-time theory then exhibits both zero and $\pi$ modes for $\eta>0$ and neither for $\eta<0$, again matching Fig.~\ref{phase_diagram} along the line $\beta=0$.

Thus we have demonstrated a one-to-one mapping between the zeroes of $i\mathcal{F}-\mathcal{F}'$ and the zeros of the Floquet Schr\"odinger operator $i\partial_t - H_F$, exactly for PBC and in the thermodynamic limit for spatial OBC. Generalizing this mapping beyond the $\beta=0$ line is the subject of the next section.

\section{Generalizing the Floquet-to-Lattice Mapping}
\label{sec: Off-axis}
We now propose a Floquet-to-lattice mapping that is valid for any $\eta$ and $\beta$, i.e., irrespective of whether or not the Floquet spectrum \eqref{theta} is $\pi$ paired.
In the previous section, solutions to Eq.~\eqref{eq:ff'p0} were automatically $\pi$ paired due to the frequency independence of $\mathcal{F}'$.
This effectively restricted our mapping to the $\beta=0$ line, where the quasienergy spectrum is $\pi$ paired.
Inspired by the temporal Wilson term construction reviewed in Sec.~\ref{sec: Lattice}, which removes $\pi$ pairing by effectively endowing the fermion mass with a frequency dependence, we will now introduce frequency dependence to the operator $\mathcal F'$ and its eigenvalues. As a result, unlike in the discussion of Sec.~\ref{sec: Floquet-to-lattice},
$\mathcal{F}'$ can no longer be interpreted as a Hamiltonian.
For example, the Wilson-Dirac operator~\eqref{gwd} cannot be expressed as $\gamma_0\partial_0+\gamma_0 H$ in the presence of a temporal Wilson term with parameter $R_0\neq0$.
This term adds higher-order time derivatives to the action and therefore affects the quantization of the theory itself.
We expand on this discussion in Appendix~\ref{app:H}.

To define the Floquet-to-lattice mapping for $\beta\neq 0$, we match solutions to Eq.~\eqref{eq:ff'p0} to solutions of the Floquet eigenvalue problem assuming that the operator $\mathcal F'$ takes the form of an $N/2$-site Wilson-Dirac model~\eqref{F'} modified to have frequency-dependent $R_1$ and $m$.
A similar mapping can be performed assuming $\mathcal F'$ is an $N$-site SSH model of the form~\eqref{ssh} with frequency-dependent $u$ and $v$---this is discussed in Appendix~\ref{app:SSH}.
In either case, the frequency-dependence of these parameters should be chosen such that $\mathcal F'$ reduces to the appropriate frequency-independent SSH~[Eq.~\eqref{ssh}] or Wilson-Dirac~[Eq.~\eqref{hwd}] Hamiltonian in the limit $\beta\to0$.

The construction of $\mathcal F'$ as a function of $\beta$ and $\eta$ is such that we avoid introducing any additional mass terms in $\mathcal{F}'$ that can break symmetries which were not broken at $\beta=0$. This is motivated by the fact that we are trying to  recover the physics of Floquet symmetry protected topological phases using discrete-time systems.
Keeping the form of $\mathcal F'$ intact as we vary $\beta$ and $\eta$ ensures that we do not change the symmetries of the discrete-time model as we move around in the Floquet parameter space. 

\subsection{PBC Eigenvalues}
We begin with the following Euclidean Wilson-Dirac-like action (assuming PBC) with frequency dependent $m$ and $R_1$
\begin{widetext}
\begin{align}
\mathcal{S}_{\text{MWD}}=\int \frac{d^2p}{(2\pi)^2}\,\,& \bar{\psi}
\left[-\frac{i}{T}\gamma^{0}\sin(p_0T)-R_1(p_0)\gamma^{1}\sin p_1+m(p_0)+R_1(p_0)(1-\cos p_1)\right]\psi.
\end{align}
The subscript MWD stands for ``modified Wilson-Dirac," referring to the frequency-dependence of the parameters. 
This corresponds to considering the operator
\begin{align}
\mathcal{F}'=\gamma_0\left[-R_1(p_0)\gamma_{1}\sin p_1+m(p_0)+R_1(p_0)(1-\cos p_1)\right],\end{align}
with eigenvalues
\beq
\label{zeta}
f'_{\pm}=\pm\sqrt{[R_1(p_0)]^2\sin^2 p_1+\left[m(p_0)+R_1(p_0)(1-\cos p_1)\right]^2}.
\eeq
We call this theory the modified Wilson-Dirac (MWD) theory. Our task is to choose frequency-dependent $m$ and $R_1$ such that the solutions of Eq.~\eqref{eq:ff'p0} with PBC match the quasienergy spectrum \eqref{theta}  of the Floquet system. To formulate the matching conditions, we follow the discussion below Eq.~\eqref{eq:epsilon_beta0} by defining two quasienergy branches,
\begin{subequations}
\label{map}
\begin{align}
\begin{split}
\epsilon_{\pm}^1(p_1)&=\epsilon_{\pm}\left(\frac{p_1}{4}+\frac{\pi}{2}\right),\\
\epsilon_{\pm}^2(p_1)&=\epsilon_{\pm}\left(\frac{p_1}{4}\right),
\end{split}
\end{align}
where $-\pi<p_1\leq \pi$, corresponding to the momentum intervals $[\pi/4,3\pi/4)$ and $[-\pi/4,\pi/4)$, respectively as seen from the argument of the RHS of the above equation. Without $\pi$ pairing, we no longer have that $\sin(T\epsilon^1_{\pm})=\sin(T\epsilon^2_{\pm})$ and must therefore separately define
\begin{align}
\begin{split}
E_{\text{MWD},\pm}^1(p_1)
&=
\frac{1}{T}
\sin\left[T \epsilon^1_{\pm}(p_1)\right],\\
E_{\text{MWD},\pm}^2(p_1)&=\frac{1}{T}\sin\left[T \epsilon^2_{\pm}(p_1)\right],
\end{split}
\end{align}
\end{subequations}
Then, we demand that the sine-transformed quasienergies $E^i_{\text{MWD},\pm}$ match the eigenvalues \eqref{zeta} of the MWD operator when evaluated at frequencies $p_0=\epsilon^i_{\pm}$, i.e.,
\beq
E_{\text{MWD},\pm}^{i}(p_1)&=&f'_{\pm}\big|_{p_0=\epsilon^i_{\pm}(p_1)}\nonumber\\
&=&\pm\sqrt{\{R_1[\epsilon_{\pm}^i(p_1)]\}^2\sin^2 p_1+\left\{m[\epsilon_{\pm}^i(p_1)]+R_1[\epsilon_{\pm}^i(p_1)](1-\cos p_1)\right\}^2}
\label{wd},
\eeq
\begin{figure*}
    \centering
    \includegraphics[width = \textwidth]{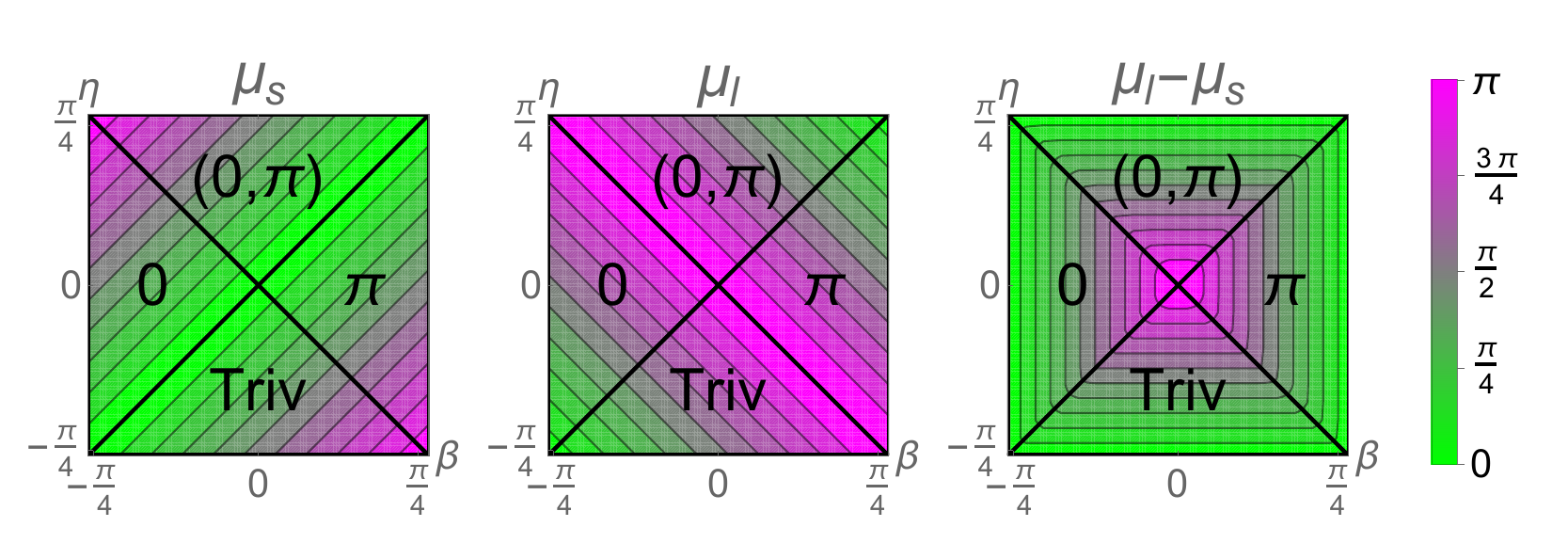}
        \caption{The smallest ($\mu_s$) and largest ($\mu_l$) eigenvalues of the Floquet Hamiltonian superimposed on the Floquet phase diagram.}
    \label{p0maxmin}
\end{figure*}
for $i=1,2$. Interestingly, using $p_0=\epsilon^i_{\pm}(p_1)$ [including inverting this relation using Eq.~\eqref{theta} to eliminate $p_1$], both of the above equations (for $i=1$ and $i=2$) reduce to
\beq
\frac{1}{T}\sin(Tp_0)=\sqrt{m^2+4 R_1(m+R_1)\left[1-\left(\frac{\cos p_0-\sin 2\beta \sin 2\eta}{\cos 2\beta \cos 2\eta}\right)^2\right]},
\eeq
\end{widetext}
where we have suppressed the $p_0$ dependence of $m$ and $R_1$ for compactness. Thus we have a single relation between $R_1(p_0)$ and $m(p_0)$ without a unique expression for either. 
To obtain unique expressions for $m(p_0)$ and $R_1(p_0)$ we need another condition. We have empirical evidence that in order to keep $m$ and $R_1$ real valued for all values of $p_0$, $\eta$ and $\beta$ we must fix
\beq
\begin{split}
2 R_1\left(R_1+m\right)&= 2 R_1\left(R_1+m\right)\big|_{\beta=0}\\
&=\frac{\cos^2(2\eta)}{2T^2}.
\end{split}
\label{imp}
\eeq
We will see, remarkably, that the same choice of parameters ensures that the OBC spectrum of the modified Wilson-Dirac theory matches one to one with the OBC Floquet spectrum.

Solving Eqs.~\eqref{wd} and \eqref{imp} allows us to extract $R_1(p_0)$ and $m(p_0)$ at all $p_0=\epsilon_{\pm}^i(p_1)$. 
For any $\eta$ and $\beta$, $|\epsilon_{\pm}^{i}(p_1)|$ takes values within a range set by the lowest and highest positive eigenvalues of the Floquet Hamiltonian $H_F$. We call the two limits $\mu_s$ and $\mu_l$. For $\eta$ and $\beta$ close to $0$, $\mu_s$ and $\mu_l$ are close to $0$ and $\pi/T$, respectively, such that $\mu_l-\mu_s\approx \frac{\pi}{T}$. However, as shown in Fig.~\ref{p0maxmin}, $\mu_s$ and $\mu_l$ move closer together as we move away from the center of the phase diagram in Fig.~\ref{phase_diagram}. Therefore if we demand that Eqs.~\eqref{wd} and~\eqref{imp} are solved only for $p_0=\epsilon_{\pm}^i(p)$, we do not have a unique functional form for $R_1(p_0)$ and $m(p_0)$ outside of the range $\mu_s<|p_0|<\mu_l$. 
Thus, in order to complete the mapping, we will
extrapolate the solution to Eqs.~\eqref{wd} and~\eqref{imp} over the full range $-\frac{\pi}{T}\leq p_0<\frac{\pi}{T}$.
The result is
\begin{subequations}
\label{MWDsol}
\begin{align}
\begin{split}
    m(p_0) &= \pm\frac{1}{T}\sqrt{\sin^2(2\eta)+\delta(p_0)},
    \\
    R_1(p_0) &= \frac{1}{2}\Big(\pm\frac{1}{T}\sqrt{1+\delta(p_0)}-m(p_0)\Big),
\end{split}
\end{align}
where 
\begin{equation}
\label{delta}
    \delta(p_0) = \bigg(\tan(2\beta)\cos(p_0T)-\frac{\sin(2\eta)}{\cos(2\beta)}\bigg)^2-\sin^2(2\eta).
\end{equation}
This allows us to rewrite $m(p_0)$ as 
\begin{equation}
    m(p_0) = \pm\frac{1}{T}\Big|\tan(2\beta)\cos(p_0 T)-\frac{\sin(2\eta)}{\cos(2\beta)}\Big|.
\end{equation}
\end{subequations}
Note that there is some freedom to choose certain signs above without modifying the PBC spectrum---we fix these signs in the next section. Note that any of the solutions \eqref{MWDsol} produce a bulk OBC spectrum that matches the PBC spectrum up to corrections $\sim 1/N$.

\subsection{OBC eigenvalues}
With Eqs.~\eqref{MWDsol} in hand, we must now choose the appropriate branches of these solutions so as to match the phase diagram of the Euclidean discrete-time theory with that of the Floquet theory.
Since the bulk PBC spectra match for any choice of signs in Eqs.~\eqref{MWDsol}, we fix these signs by matching the boundary spectra of the two models everywhere in the Floquet phase diagram of Fig.~\ref{phase_diagram}.
The boundary spectrum of the Floquet theory with OBC contains either a zero-quasienergy mode or a $\pi/T$-quasienergy mode. In order to establish a mapping to the discrete-time theory, we thus will 
only need to concern ourselves with zero- and $\pi/T$-frequency boundary modes of the lattice fermion operator $\mathcal{F}+\mathcal{F}'$. Here, the phrase ``boundary mode" denotes zero-eigenvalue eigenstates of the full Euclidean fermion operator that are localized on the boundary.
Since we want to identify zero modes of $\mathcal{F}+\mathcal{F}'$ that are also zero modes of the operator $\mathcal{F}=\nabla_0$, they must also be zero modes of $\mathcal{F}'$ (with OBC). In this case, the states of interest are indeed poles of the Euclidean fermion propagator.

We first consider the eigenvalues of $\mathcal{F}'$ with OBC inside the region $|\beta|<|\eta|$, which corresponds to the $(0,\pi)$ phase for $\eta>0$ and the trivial phase for $\eta<0$.
In order for $m(p_0)$ and $R_1(p_0)$ to match to their expressions along $\beta=0$ as given in Eq.~\eqref{m} we must pick the ``$+$" branch for $R_1$,
\begin{align}
    R_1(p_0) &= \frac{1}{2}\Big(\frac{1}{T}\sqrt{1+\delta(p_0)}-m(p_0)\Big).
    \label{Rn}
\end{align}
\begin{figure}
    \centering
    \includegraphics[width = \columnwidth]{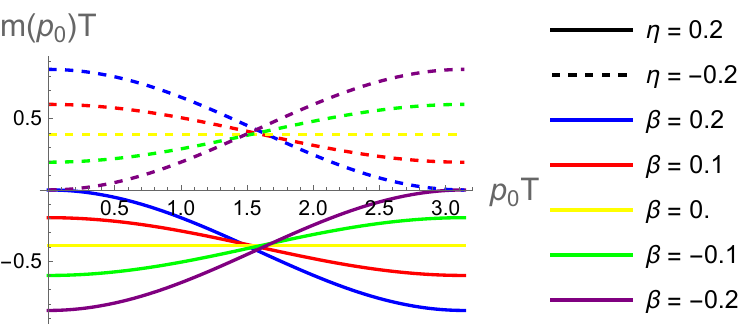}
    \caption{The frequency dependent mass $m(p_0)$ as given by equations \eqref{eta>0} and \eqref{eta<0}, plotted for a few different values of $\beta$ and $\eta$. }
    \label{m_and_R_illustration}
\end{figure}
This follows from observing that $\delta(p_0)=0$ when $\beta=0$ [see Eq.~\eqref{delta}].
Then, recalling from Sec.~\ref{sec: Lattice} that boundary modes appear in the Wilson-Dirac model when $m<0$, we demand that $m(p_0)$ is positive for $\eta<0$ (i.e., in the trivial phase) and negative for $\eta>0$ (i.e., in the $(0,\pi)$ phase).
This implies that we make the unique choices
\begin{align}
    m(p_0) &= -\frac{1}{T}\Big|\tan(2\beta)\cos(p_0 T)-\frac{\sin(2\eta)}{\cos(2\beta)}\Big|,
    \label{eta>0}
\end{align}
for drive parameters $(\eta>0, -\eta<\beta<\eta)$, and
\begin{equation}
    m(p_0) = \frac{1}{T}\Big|\tan(2\beta)\cos(p_0T)-\frac{\sin(2\eta)}{\cos(2\beta)}\Big|,\label{eta<0}
\end{equation}
for drive parameters $(\eta<0, \eta<\beta<-\eta)$. In Fig.~\ref{m_and_R_illustration}, we plot $m$ as a function of $p_0$ for $\eta = \pm 0.2$ and several values of $\beta$. Notably, the mass is either positive or negative definite depending on the sign of $\eta$, except at $\beta = \pm \eta$ where $m=0$ at $p_0T=0$ or $\pi$. These zeros of the mass correspond to quasienergy gap closings around $\epsilon=0$ and $\pi/T$, respectively. Furthermore, note that both $m$ and $m^2$ are smooth functions of $p_0$ for any $\eta$ and $\beta$, so long as $|\beta|\leq|\eta|$.

\begin{figure}
    \centering
    \includegraphics[width=\columnwidth]{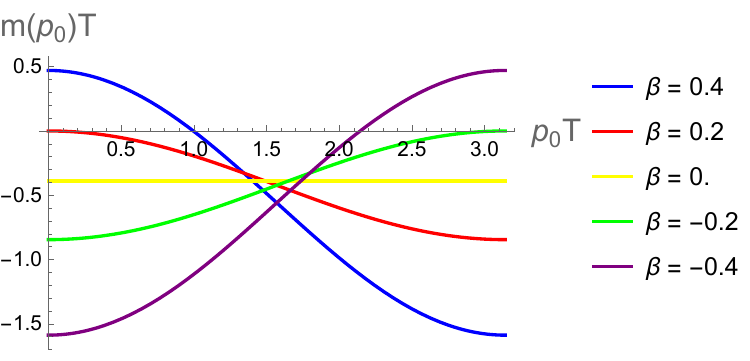}
    \caption{The frequency dependent mass $m(p_0)$ as given by \eqref{ch1} and \eqref{ch2}, computed using $\eta=0.2$.}
    \label{m_and_R_illustration_2}
\end{figure}

Next, we turn to the opposite regime $|\beta| >|\eta|$, which corresponds to the $\pi$ phase for $\beta>0$ and the $0$ phase for $\beta<0$. In this regime, for any choice of the sign of $m(p_0)$ in Eq.~\eqref{MWDsol}, there is some $p_0=p_0'$ with $\pi/T>p_0'>0$ such that $m(p_0')=0$. To ensure that $m$ is a smooth function of $p_0$, we must therefore demand that the sign of $m$ change at $p_0'$. To fix the sign on either side of $p_0'$, we again appeal to the fact that boundary modes appear when $m<0$. Thus, in the $0$ phase ($\beta < 0$), we demand that $m<0$ at $p_0=0$ and $m>0$ at $p_0=\pi/T$, such that
\begin{align}
m(p_0)\!&=\!\frac{1}{T}\Big|\tan(2\beta)\cos(p_0 T)\!-\! \frac{\sin(2\eta)}{\cos(2\beta)}\Big|\, \text{sgn}(p_0\!-\!p_0'),
\label{ch1}
\end{align}
for $\beta < \eta < -\beta$. Similarly, in the $\pi$ phase ($\beta>0$), we demand that $m<0$ at $p_0=\pi/T$ and $m>0$ at $p_0=0$, such that
\begin{align}
m(p_0)\!&=\!-\frac{1}{T}\Big|\tan(2\beta)\cos(p_0 T)\!-\! \frac{\sin(2\eta)}{\cos(2\beta)}\Big|\, \text{sgn}(p_0\!-\!p_0'),
\label{ch2}
\end{align}
In Fig.~\ref{m_and_R_illustration_2}, we plot these expressions for $\eta=0.2$ and a few positive and negative $\beta$ values. Importantly, we see that $m$ is again a smooth function of $p_0$ and that the curves at the gapless points $\beta = \pm \eta$ match their counterparts in Fig.~\ref{m_and_R_illustration}.

\begin{figure*}
    \centering
    \includegraphics[width=\textwidth]{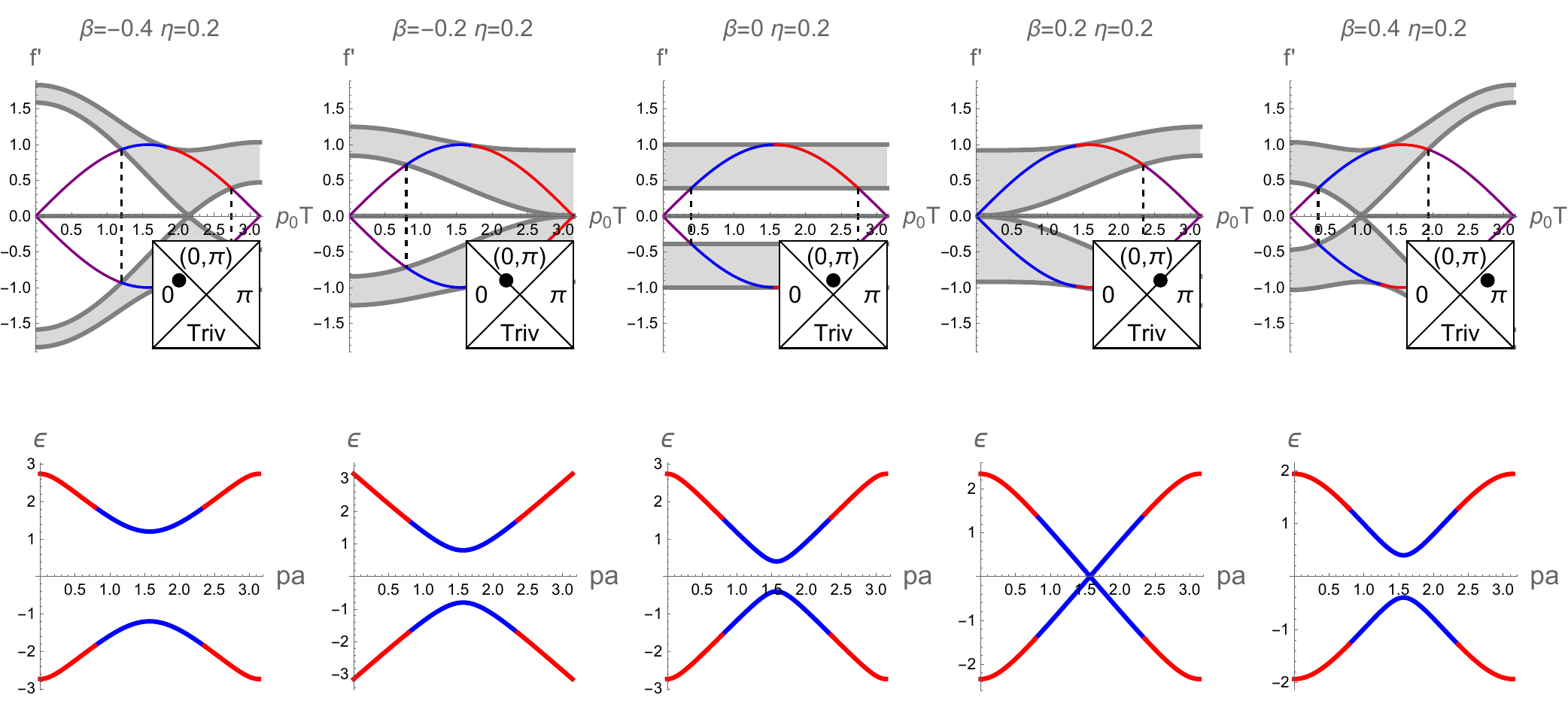}
    \caption{
    The OBC eigenvalues $f'$ of $\mathcal{F}'(m,R_1)$ (grey region surrounded by dark grey lines, top row) and the PBC spectrum of the Floquet operator $\epsilon$ (bottom row) plotted for different values of $\beta$ along the line $\eta = 0.2$. The solutions for $m(p_0)$ and $R(p_0)$ are chosen according to equation \eqref{mR_analytic}. The purple (blue and red) line represents $\pm\sin p_0 T$ for values of $p_0$ that do not (do) satisfy Eq.~\eqref{map}. Identifying these $p_0$ values with quasienergies yields the corresponding blue and red eigenvalues depicted in the bottom plot (up to finite size effects relating the bulk PBC and OBC spectra). The blue and red colors label the two quasienergy branches $i=1$ and $2$, respectively, that appear in Eqs.~\eqref{map}} 
    \label{F_horizontal}
\end{figure*}

\begin{figure*}
    \centering
    \includegraphics[width=\textwidth]{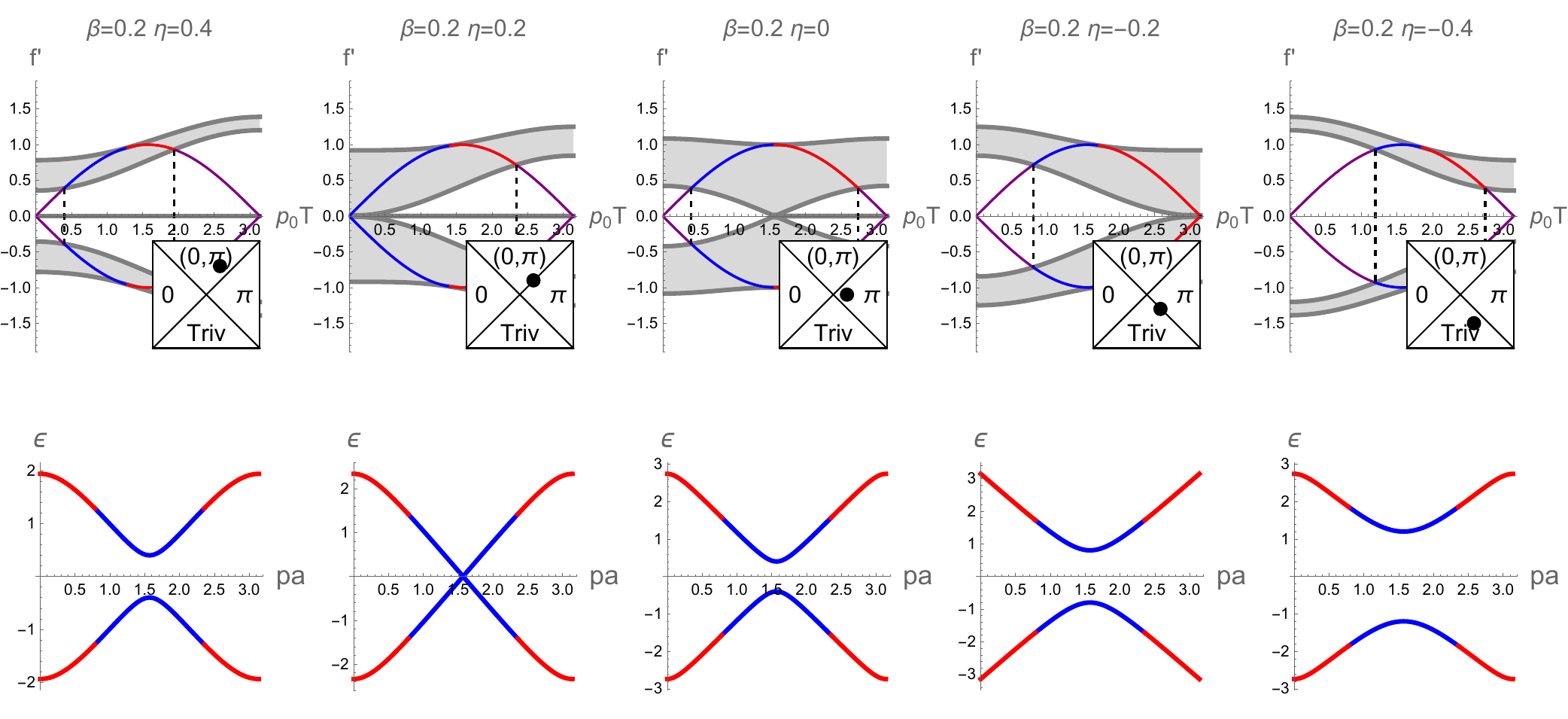}
    \caption{Same as Fig.~\ref{F_horizontal}, but instead we have fixed $\beta = 0.2$ and vary $\eta$.}
    \label{F_vertical}
\end{figure*}

Having chosen the appropriate branch for each phase in the Floquet phase diagram, it is interesting to note that they can all be summarized by the following expressions:
\begin{align}
\begin{split}
    m(p_0) &= \frac{1}{T}\Big[\tan(2\beta)\cos(p_0 T)-\frac{\sin(2\eta)}{\cos(2\beta)}\Big],
    \\
    R_1(p_0) &= \frac{1}{2}\Big(\frac{1}{T}\sqrt{1+\delta(p_0)}-m(p_0)\Big).
    \label{mR_analytic}
\end{split}
\end{align}
In other words, $m(p_0)$ and $R_1(p_0)$ are both analytic and monotonic functions of $p_0$ everywhere in the Floquet phase diagram. 
In Figs.~\ref{F_horizontal} and~\ref{F_vertical}, we have used these expressions to compute the eigenvalues of the operator $\mathcal{F}'$ with OBC for a few values of $\beta$ and $\eta$ along a horizontal and a vertical cut through the Floquet phase diagram, respectively.
For the computation we have used $2N=200$ lattice points.
The OBC eigenvalues are visualized as filled gray bands between the extremal eigenvalues, which are plotted as dark gray lines. For values of $\eta$ and $\beta$ corresponding to the 0 and $\pi$ phases, a flat band of zero-eigenvalues connecting to $p_0=0$ and $\pi/T$, respectively, indicates the presence of zero or $\pi$ boundary modes. For $\eta$ and $\beta$ corresponding to the $(0,\pi)$ phase, a flat band of zeroes connecting $p_0=0$ and $\pi/T$ indicates the presence of both $0$ and $\pi$ boundary modes. Furthermore, to visually represent the correspondence between bulk OBC eigenvalues and quasienergies via the spectral mapping~\eqref{map}, we superimpose a plot of $\pm \sin(p_0T)$. The regions where the bulk eigenvalues do no intersect this line, the mapping \eqref{map} cannot be satisfied, and we color this part of the line purple. The values of $p_0$ for which $\pm \sin(p_0T)$ does intersects the gray eigenvalue bands, are identified with the quasienergies shown below each of the $\mathcal{F}'$ eigenvalue plots. The blue and red colors of the quasienergies and $\pm\sin(p_0 T)$ correspond to the branches 1 and 2 in Eqs.~\eqref{map}, respectively. Note that as the eigenvalues $f'$ are symmetric in $p_0$, we have chosen to plot only positive $p_0$ in the upper panels. Thus the highlighted $p_0$ values correspond to the positive quasienergies in the lower panels. For plots representing the spectrum along the gapless lines $|\eta|=|\beta|$, note that the band of highlighted $p_0$ values touches $p_0=0$ or $\pi/T$, indicating quasienergy gap closings at quasienergy $0$ or $\pi/T$.

\begin{figure*}[ht!]
    \centering
    \includegraphics[width = \textwidth]{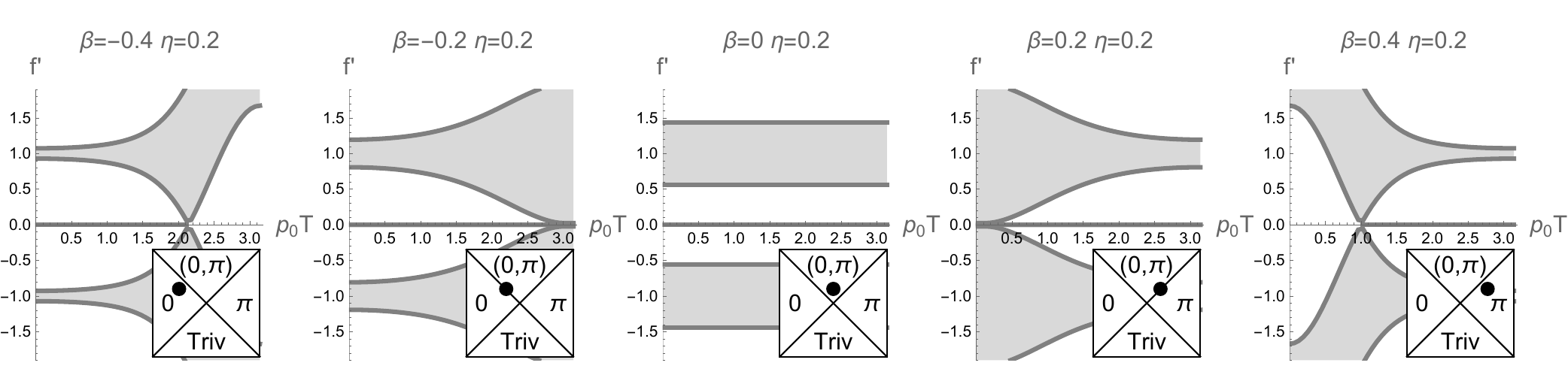}
    \caption{The eigenvalues $f'$ of $\mathcal{F}'$, depicted as in Figs.~\ref{F_horizontal} and \ref{F_vertical}, after the substitution $m(p_0)\to\frac{m(p_0)}{R_1(p_0)}$ and $R_1(p_0)\to1$.}
    \label{m_over_R}
\end{figure*}

\subsection{Frequency dependence near $\beta,\eta\to 0$ and comparison to standard Wilson-Dirac fermions}

We now make contact with the discussion in Sec.~\ref{sec: Lattice} by examining the expressions for $m(p_0)$ and $R_1(p_0)$ close to the center of the Floquet phase diagram (i.e., in the limit $\beta,\eta\to 0$).
Expanding to leading order in $\beta$ and $\eta$, Eq.~\eqref{mR_analytic} reduces to
\begin{align}
\begin{split}
    m(p_0) &= -\frac{2\eta}{T}+\frac{2\beta}{T}\cos(p_0 T),
    \\
    R_1(p_0) & = \frac{1}{T}\Big[\frac{1}{2}+2\eta-2\beta\cos(p_0 T)\Big].
    \label{mr}
\end{split}
\end{align}
These expressions capture the essence of the entire phase diagram: since all four phases meet at $\eta=\beta=0$, all four phases are present for small $\eta,\beta$. 
For example, the expression for $m(p_0)$ in Eq.~\eqref{mr} clearly shows gap closures for $p_0=0$ at $\beta=\eta$ and for $p_0=\frac{\pi}{T}$ for $\beta=-\eta$. Similarly, one can verify that the boundary mode behavior is also consistent with the Floquet phase diagram.

One of the benefits of considering the frequency dependence for small $\eta,\beta$ is that the expressions for $m$ and $R_1$ reduce to linear functions of $\cos(p_0 T)$. Fourier transforming yields the following functions of Euclidean time $x_0$ (again up to leading order in $\beta$ and $\eta$):
\beq
\begin{split}
m(x_0)&=-\frac{2(\eta-\beta)}{T}+\beta T\,\nabla_0^2,\\
R_1(x_0)&=\frac{1}{T}\left(\frac{1}{2}+2(\eta-\beta)-\beta T^2\nabla_0^2\right).
\end{split}
\eeq
More generally, the Euclidean-time structure of the added terms can be obtained by performing a series expansion of Eq.~\eqref{mR_analytic} in small $\beta$ and $\eta$, followed by the substitution
\begin{align}
\frac{1}{T^2}\cos(p_0 T) \to \frac{1}{2}\nabla_{0}^2+\frac{1}{T^2}.
\end{align}
The result is an action with higher-order time derivatives, which therefore remains local in Euclidean time for small $\eta$ and $\beta$.

The main difference between the modified Wilson-Dirac parameters of Eq.~\eqref{mr} and the standard Wilson-Dirac parameters lies in the behavior of $R_1$, which is typically frequency independent and often set to 1. It is clear that setting $R_1=1$ would not satisfy the spectral matching conditions \eqref{map}.
However, if we demand that the discrete time theory only replicate the gap closures of the Floquet theory and its boundary modes, we mayconsider a lattice field theory defined by the eigenvalue problem
\begin{align}
\sin(p_0 T)\ket{\psi(p_0,p_1)}=\frac{1}{\kappa(p_0)}\mathcal{F}'(p_0, p_1)T\ket{\psi(p_0,p_1)},
\end{align}
which can be viewed as a replacement of Eq.~\eqref{eq:ff'p0}. Here, $\kappa(p_0)$ is some arbitrary function of $p_0$ which does not have zeroes. 
Now, we can set $\kappa(p_0)=R_1(p_0)$ since $R_1$ does not have zeroes except at the boundaries of the phase diagram at $\eta=\pm\pi T/4$. The corresponding Euclidean action has the form of a Wilson Dirac action with unit Wilson parameter, given by 
\begin{align}
\begin{split}
\mathcal{S}_{\text{WD}}=\int \frac{d^2p}{(2\pi)^2}&\,\, \bar{\psi}
\big[-i\gamma_{0}\sin (p_0T)-i \gamma_{1}\sin p_1
\\
&+m'(p_0)+(1-\cos p_1)\big]\psi
\end{split},
\end{align}
where $m'(p_0)=m(p_0)/R_1(p_0)$. For small $\eta$ and $\beta$ we can write
\begin{align}
    m'(p_0)\approx& -\frac{4}{T}[\eta-\beta\cos(p_0 T)+\dots],
    \nonumber
    \\
    &\implies m'(x_0)=\frac{4(\beta-\eta)}{T}+2\beta \,T\nabla_0^2+\cdots
\end{align}
As can be seen from the eigenvalue plots in Fig.~\ref{m_over_R}, the resulting bulk theory goes gapless in the exact same places as the original one, while also having an identical boundary mode spectrum.

\section{Conclusion}
In this paper we answer, at least in one model, a conceptual question: {\it Can periodically driven quantum systems like Floquet insulators be reinterpreted as discrete-time theories without any drive?} We raised this question in Ref.~\cite{FloquetLattice1} and answered it in the affirmative for a $1+1$ dimensional Floquet insulator model in a parameter regime where the Floquet spectrum exhibits $\pi$ pairing. We demonstrated that the Floquet spectrum can be reproduced via Fermion doubling in a discrete time theory on a smaller spatial lattice (with either half or one-quarter of the sites of the original model, depending on whether the fermions are spinless or spinful). This reinterpretation relied on the fact that a symmetric discrete-time derivative maps to $\frac{1}{T}\sin(p_0 T)$ in Fourier space. Therefore when the relevant Schr\"odinger equation is discretized, we automatically gain a pair of solutions that are $\pi$-paired. However, this approach breaks down away from the $\beta=0$ line in Fig.~\ref{phase_diagram}, where $\pi$ pairing is absent.
In this paper we extend our approach and show that the entire phase diagram of the Floquet model can be reinterpreted as a discrete-time theory with an undriven Hamiltonian.  To formulate such a reinterpretation in the absence of $\pi$ pairing, we took inspiration from the temporal Wilson term in lattice field theory, which breaks $\pi$ pairing by adding a frequency dependence to the mass term. We constructed an explicit mapping from the Floquet spectrum to a discrete-time Wilson-Dirac-like theory with frequency dependent mass and spatial Wilson parameter. A similar mapping to an SSH-like model with frequency dependent hoppings is given in Appendix~\ref{app:SSH} The mapping is exact for PBC and automatically holds for the bulk spectrum in OBC up to corrections that go as $1/N$; the low-lying boundary spectra match exactly by construction. Remarkably the theory on the lattice field theory side is completely local in space. 

There are several questions that remain to be explored. The correspondence between the Floquet spectrum and discrete-time theories that we constructed here apply to free theories on either side. The question then naturally arises of whether such a correspondence can be extended to interacting theories. More specifically, interacting Floquet systems can heat up to infinite temperature after a prethermal timescale, destabilizing the Floquet spectrum~\cite{BukovReview,D'Alessio,Abanin17}. It will be interesting to examine if this phenomenon can be related to the destabilization of topological phases on the discrete time side. Another direction of research involves extending the correspondence found here to higher-dimensional Floquet systems \cite{Rudner13} and lattice fermion theories. This may help shed light on whether there exist any ties between the bulk-boundary correspondences in Floquet systems~\cite{Kitagawa,Rudner13,Nathan,Carpentier,Fruchart} and lattice field theories \cite{Golterman:1992ub}. 
Finally, it is important to note that there have been several proposals to simulate different types of interactions using Floquet systems 
\cite{PhysRevB.95.024431, PRXQuantum.3.020303, PhysRevX.10.031002,Wintersperger_2020, Ciavarella:2022tvc}. Similar efforts have also been made to simulate gauge theory Hamiltonians using Floquet systems \cite{Schweizer_2019,PhysRevX.4.031027,PhysRevLett.111.185301,RevModPhys.89.011004}. Our findings may illuminate new ways to add a fermionic sector to these target theories. 
\label{sec: Conclusion}

\begin{acknowledgments}
T.I.~acknowledges support from the National Science Foundation under Grant No.~DMR-2143635. S.S.~and L.S.~acknowledge support from the U.S. Department of Energy,
Nuclear Physics Quantum Horizons program through the
Early Career Award DE-SC0021892.
\end{acknowledgments}

\begin{appendix}
\section{Two solutions for each frequency: An example}
\label{app:eigen}
In Sec.~\ref{sec: Floquet-to-lattice} it is stated that, for every crystal momentum $p_1$, Eq.~\eqref{eq:ff'p0} is satisfied by two different frequencies $p_0>0$ (and also two different $p_0<0$) irrespective of the details of $\mathcal{F}'$. As an example, when $\mathcal{F}'$ is given by \eqref{F'}, we have
\begin{widetext}
\begin{equation}
\label{eq:twosol}
p_0 T = \begin{cases}+\arccos\Big[-\frac{R_0\pm\sqrt{R_0^2+4(R_0^2-1)[2m+R_0+R_0^2-1+R_1(1-\cos p_1)(1+R_1+R_1\cos p_1)]}}{2(R_0^2-1)}\Big]\\
-\arccos\Big[-\frac{R_0\pm\sqrt{R_0^2+4(R_0^2-1)[2m+R_0+R_0^2-1+R_1(1-\cos p_1)(1+R_1+R_1\cos p_1)]}}{2(R_0^2-1)}\Big]
\end{cases}.
\end{equation}
\end{widetext}
Each line on the RHS of Eq.~\eqref{eq:twosol} corresponds to two different frequencies with one of the lines producing two positive frequencies and the other producing two negative ones.
In the case where $\mathcal F'$ is frequency-independent (i.e., where $R_0=0$), the two positive frequencies become $\pi$ paired, i.e.~they sum to $\pi/T$. Similarly the negative ones sum to $-\pi/T$.
\section{Hamiltonian}
\label{app:H}

As mentioned at the beginning of Sec.~\ref{sec: Off-axis}, $\mathcal{F}'$ cannot be interpreted as a Hamiltonian when its parameters are frequency-dependent. To elaborate, consider the standard Wilson-Dirac action in two spacetime dimensions continued to Minkowski space, 
\beq
\mathcal S=\int d^2x\,\,\bar{\psi}\left(i\gamma^{\mu}\partial_{\mu}-m-\frac{R_{\mu}}{2}\partial_{\mu}^2\right)\psi .
\eeq
Here the $\gamma$ matrices satisfy $\{\gamma^{\mu},\gamma^{\nu}\}=2\eta^{\mu\nu}$ with $\eta$ being the Minkowski metric ($\eta^{00}=1, \eta^{11}=-1$). We have also allowed the Wilson parameters to be different in the space and time directions. 
Clearly, $\mathcal{F}'$ here is frequency dependent. We observe that the conjugate momentum to $\dot{\psi}\equiv\partial_0\psi$ is
\beq
\pi=i\bar{\psi}\gamma^0+\frac{R_0}{2}\dot{\bar{\psi}}
\eeq
which leads to the Hamiltonian
\beq
H'=m\bar{\psi}\psi-\frac{R_1}{2}\bar{\psi}\partial_1^2\psi-i R_1\bar{\psi}\gamma^1\partial_1\psi.
\eeq
The one-body Hamiltonian operator extracted from $H'$, i.e.,
\begin{align}
h=m\gamma^0-\frac{R_1}{2}\gamma^0\nabla_1^2-iR_1\gamma^0\gamma^1\partial_1,
\end{align}
is time-independent and in fact identical to what one would obtain if $R_0=0$. The difference between the lattice spectra of the two cases $R_0=0$ and $R_0\neq 0$ arises from the fact that, on the one hand, for $R_0=0$ the Minkowski equation of motion reduces to the Schr\"odinger equation with the Hamiltonian $h$. On the other hand, for $R_0\neq 0$ the Minkowski equation of motion is not a Schr\"odinger equation, even though the Hamiltonian operator, $h$, remains unchanged.

\section{Mapping to an SSH-like model}
\label{app:SSH}

In Sec.~\ref{sec: Off-axis} we focus on a spectral mapping between the Floquet model \eqref{u} and a modified Wilson-Dirac model in two-dimensional discrete Euclidean spacetime. Here we demonstrate that a similar mapping can be made to a modified discrete-spacetime SSH model with $\mathcal{F}'$ set to
 $\frac{v}{2} H_0+\frac{u}{2} H_1$, and where 
$u$ and $v$ are now frequency dependent. The eigenvalues of $\mathcal{F}'$ with PBC in this case are given by
\begin{equation}
 f'_{\pm}   =\pm\sqrt{u(p_0)^2+v(p_0)^2+2u(p_0)v(p_0)\cos(2p_1)},
 \vspace{10pt}
\end{equation}
for $0\leq p < \pi$.
To construct the map between the PBC Floquet spectrum and the discrete-time theory, we define
\beq
E_{\text{SSH},\pm}^1(p_1)&=&\frac{1}{T}\sin\left[T \epsilon_{\pm}\left(\frac{p_1}{2}+\frac{\pi}{4}\right)\right],\nonumber\\
E_{\text{SSH},\pm}^2(p_1)&=&\frac{1}{T}\sin\left[T \epsilon_{\pm}\left(\frac{p_1}{2}-\frac{\pi}{4}\right)\right],\nonumber\\
\tilde{\epsilon}_{\pm}^1(p_1)&=&\frac{1}{T}\sin^{-1}[T E_{\text{SSH},\pm}^1(p_1)]=\epsilon_{\pm}\left(\frac{p_1}{2}+\frac{\pi}{4}\right),\nonumber\\
\tilde{\epsilon}_{\pm}^2(p_1)&=&\frac{1}{T}\sin^{-1}[T E_{\text{SSH},\pm}^2(p)]=\epsilon_{\pm}\left(\frac{p_1}{2}-\frac{\pi}{4}\right),\nonumber\\
\eeq
and impose as before
\begin{widetext}
\beq
E_{\text{SSH},\pm}^{i}(p_1)&=&
f'_{\pm}\big|_{p_0=\tilde{\epsilon}_{\pm}^i(p_1)}\nonumber\\
&=&\sqrt{[u(\tilde{\epsilon}_\pm^i(p_1))]^2+[v(\tilde{\epsilon}_\pm^i(p_1))]^2+2[u(\tilde{\epsilon}_\pm^i(p_1))][v(\tilde{\epsilon}_\pm^i(p_1))]\cos(2p_1)},
\label{first}
\eeq
\end{widetext}
where $i=1, 2$ label the two quasienergy branches.
Solving this equation we get a relation between $u(p_0)$ and $v(p_0)$. However, just as before, this relation does not uniquely fix $u(p_0)$ and $v(p_0)$. Again, we set 
\beq
u(p_0)v(p_0)=\frac{1-\sin^2(2\eta)}{2T^2},
\eeq
to keep $u$ and $v$ real. With this we get the following solutions:
\begin{align}
    u(p_0) &= \frac{\sqrt{1+\delta(p_0)}\pm\Big|\tan(2\beta)\cos(p_0 T)-\frac{\sin(2\eta)}{\cos(2\beta)}\Big|}{2T},
    \nonumber
    \\
    v(p_0) &= \frac{1}{T}-u(p_0).
\end{align}
The next step is then to identify the correct solution branches corresponding to the different parts of the Floquet phase diagram. We first consider the region $-\eta<\beta<\eta$ for $\eta>0$. Here we pick the branch 
\begin{equation}
    u(p_0) = \frac{\sqrt{1+\delta(p_0)}+\Big|\tan(2\beta)\cos(p_0 T)-\frac{\sin(2\eta)}{\cos(2\beta)}\Big|}{2T},
\end{equation}
so as to match onto our solution \eqref{uv} along $\beta=0, \eta>0$. By the same argument, when $\eta<0$ and $-\eta<\beta<\eta$, we must choose 
\begin{equation}
    u(p_0) = \frac{\sqrt{1+\delta(p_0)}-\Big|\tan(2\beta)\cos(p_0 T)-\frac{\sin(2\eta)}{\cos(2\beta)}\Big|}{2T}\label{u eta<0}.
\end{equation}
Note that the spectrum of $\mathcal{F}'$ with PBC is gapped as long as $u(p_0)\neq v(p_0)$, meaning that the gap closes when $\Big|\tan(2\beta)\cos(p_0 T)-\frac{\sin(2\eta)}{\cos(2\beta)}\Big|=0$. At this point we can demand that the OBC spectrum of this lattice theory match with the Floquet phase diagram to get
\begin{align}
    u(p_0) &= \frac{\sqrt{1+\delta(p_0)}+\tan(2\beta)\cos(p_0 T)-\frac{\sin(2\eta)}{\cos(2\beta)}}{2T},
    \nonumber
    \\
    v(p_0) &= \frac{1}{T}-u(p_0).
\end{align}
This expression holds for all $\eta$ and $\beta$.  

As in the Wilson-Dirac case, it is instructive to consider the expressions for $u(p_0)$ and $v(p_0)$ around the center of the Floquet phase diagram, i.e.~for $\eta,\beta\rightarrow 0$. Retaining only the first order terms in $\eta$ and $\beta$ we find
\begin{align}
\begin{split}
    u(p_0) &= \frac{1}{2T}[1-2\eta+2\beta\cos(p_0 T)],
    \\
    v(p_0) &= \frac{1}{2T}[1+2\eta-2\beta\cos(p_0 T)].
\end{split}
\end{align}
Converting these expressions into an expression for the mass, we obtain
\begin{align}
\begin{split}
    m(p_0) &\propto u(p_0)-v(p_0)\\
    &=-\frac{2\eta}{T}+\frac{2\beta}{T}\cos(p_0T),
\end{split}
\end{align}
which matches the expression for the mass given in Eq.~\eqref{mr} for the modified Wilson-Dirac model. Thus, we see an analogous connection between this model and a ``standard" Wilson-Dirac-type picture in this limit.

\end{appendix}

\bibliographystyle{apsrev4-2}
\bibliography{floquet}

\begin{thebibliography}{48}%
\makeatletter
\providecommand \@ifxundefined [1]{%
 \@ifx{#1\undefined}
}%
\providecommand \@ifnum [1]{%
 \ifnum #1\expandafter \@firstoftwo
 \else \expandafter \@secondoftwo
 \fi
}%
\providecommand \@ifx [1]{%
 \ifx #1\expandafter \@firstoftwo
 \else \expandafter \@secondoftwo
 \fi
}%
\providecommand \natexlab [1]{#1}%
\providecommand \enquote  [1]{``#1''}%
\providecommand \bibnamefont  [1]{#1}%
\providecommand \bibfnamefont [1]{#1}%
\providecommand \citenamefont [1]{#1}%
\providecommand \href@noop [0]{\@secondoftwo}%
\providecommand \href [0]{\begingroup \@sanitize@url \@href}%
\providecommand \@href[1]{\@@startlink{#1}\@@href}%
\providecommand \@@href[1]{\endgroup#1\@@endlink}%
\providecommand \@sanitize@url [0]{\catcode `\\12\catcode `\$12\catcode
  `\&12\catcode `\#12\catcode `\^12\catcode `\_12\catcode `\%12\relax}%
\providecommand \@@startlink[1]{}%
\providecommand \@@endlink[0]{}%
\providecommand \url  [0]{\begingroup\@sanitize@url \@url }%
\providecommand \@url [1]{\endgroup\@href {#1}{\urlprefix }}%
\providecommand \urlprefix  [0]{URL }%
\providecommand \Eprint [0]{\href }%
\providecommand \doibase [0]{http://dx.doi.org/}%
\providecommand \selectlanguage [0]{\@gobble}%
\providecommand \bibinfo  [0]{\@secondoftwo}%
\providecommand \bibfield  [0]{\@secondoftwo}%
\providecommand \translation [1]{[#1]}%
\providecommand \BibitemOpen [0]{}%
\providecommand \bibitemStop [0]{}%
\providecommand \bibitemNoStop [0]{.\EOS\space}%
\providecommand \EOS [0]{\spacefactor3000\relax}%
\providecommand \BibitemShut  [1]{\csname bibitem#1\endcsname}%
\let\auto@bib@innerbib\@empty
\bibitem [{\citenamefont {Bukov}\ \emph {et~al.}(2015)\citenamefont {Bukov},
  \citenamefont {D{\textquotesingle}Alessio},\ and\ \citenamefont
  {Polkovnikov}}]{BukovReview}%
  \BibitemOpen
  \bibfield  {author} {\bibinfo {author} {\bibfnamefont {M.}~\bibnamefont
  {Bukov}}, \bibinfo {author} {\bibfnamefont {L.}~\bibnamefont
  {D{\textquotesingle}Alessio}}, \ and\ \bibinfo {author} {\bibfnamefont
  {A.}~\bibnamefont {Polkovnikov}},\ }\href {\doibase
  10.1080/00018732.2015.1055918} {\bibfield  {journal} {\bibinfo  {journal}
  {Advances in Physics}\ }\textbf {\bibinfo {volume} {64}},\ \bibinfo {pages}
  {139} (\bibinfo {year} {2015})}\BibitemShut {NoStop}%
\bibitem [{\citenamefont {Cayssol}\ \emph {et~al.}(2013)\citenamefont
  {Cayssol}, \citenamefont {D{\'{o}}ra}, \citenamefont {Simon},\ and\
  \citenamefont {Moessner}}]{CayssolReview}%
  \BibitemOpen
  \bibfield  {author} {\bibinfo {author} {\bibfnamefont {J.}~\bibnamefont
  {Cayssol}}, \bibinfo {author} {\bibfnamefont {B.}~\bibnamefont {D{\'{o}}ra}},
  \bibinfo {author} {\bibfnamefont {F.}~\bibnamefont {Simon}}, \ and\ \bibinfo
  {author} {\bibfnamefont {R.}~\bibnamefont {Moessner}},\ }\href {\doibase
  10.1002/pssr.201206451} {\bibfield  {journal} {\bibinfo  {journal} {physica
  status solidi ({RRL}) - Rapid Research Letters}\ }\textbf {\bibinfo {volume}
  {7}},\ \bibinfo {pages} {101} (\bibinfo {year} {2013})}\BibitemShut {NoStop}%
\bibitem [{\citenamefont {Rudner}\ and\ \citenamefont
  {Lindner}(2020)}]{RudnerReview}%
  \BibitemOpen
  \bibfield  {author} {\bibinfo {author} {\bibfnamefont {M.~S.}\ \bibnamefont
  {Rudner}}\ and\ \bibinfo {author} {\bibfnamefont {N.~H.}\ \bibnamefont
  {Lindner}},\ }\href {\doibase 10.1038/s42254-020-0170-z} {\bibfield
  {journal} {\bibinfo  {journal} {Nature Reviews Physics}\ }\textbf {\bibinfo
  {volume} {2}},\ \bibinfo {pages} {229} (\bibinfo {year} {2020})}\BibitemShut
  {NoStop}%
\bibitem [{\citenamefont {Else}\ \emph {et~al.}(2020)\citenamefont {Else},
  \citenamefont {Monroe}, \citenamefont {Nayak},\ and\ \citenamefont
  {Yao}}]{ElseReview}%
  \BibitemOpen
  \bibfield  {author} {\bibinfo {author} {\bibfnamefont {D.~V.}\ \bibnamefont
  {Else}}, \bibinfo {author} {\bibfnamefont {C.}~\bibnamefont {Monroe}},
  \bibinfo {author} {\bibfnamefont {C.}~\bibnamefont {Nayak}}, \ and\ \bibinfo
  {author} {\bibfnamefont {N.~Y.}\ \bibnamefont {Yao}},\ }\href {\doibase
  10.1146/annurev-conmatphys-031119-050658} {\bibfield  {journal} {\bibinfo
  {journal} {Annual Review of Condensed Matter Physics}\ }\textbf {\bibinfo
  {volume} {11}},\ \bibinfo {pages} {467} (\bibinfo {year} {2020})}\BibitemShut
  {NoStop}%
\bibitem [{\citenamefont {Sacha}\ and\ \citenamefont
  {Zakrzewski}(2017)}]{SachaReview}%
  \BibitemOpen
  \bibfield  {author} {\bibinfo {author} {\bibfnamefont {K.}~\bibnamefont
  {Sacha}}\ and\ \bibinfo {author} {\bibfnamefont {J.}~\bibnamefont
  {Zakrzewski}},\ }\href {\doibase 10.1088/1361-6633/aa8b38} {\bibfield
  {journal} {\bibinfo  {journal} {Reports on Progress in Physics}\ }\textbf
  {\bibinfo {volume} {81}},\ \bibinfo {pages} {016401} (\bibinfo {year}
  {2017})}\BibitemShut {NoStop}%
\bibitem [{\citenamefont {Khemani}\ \emph {et~al.}(2019)\citenamefont
  {Khemani}, \citenamefont {Moessner},\ and\ \citenamefont
  {Sondhi}}]{KhemaniReview}%
  \BibitemOpen
  \bibfield  {author} {\bibinfo {author} {\bibfnamefont {V.}~\bibnamefont
  {Khemani}}, \bibinfo {author} {\bibfnamefont {R.}~\bibnamefont {Moessner}}, \
  and\ \bibinfo {author} {\bibfnamefont {S.~L.}\ \bibnamefont {Sondhi}},\
  }\href {\doibase 10.48550/ARXIV.1910.10745} {\enquote {\bibinfo {title} {A
  brief history of time crystals},}\ } (\bibinfo {year} {2019})\BibitemShut
  {NoStop}%
\bibitem [{\citenamefont {Hasan}\ and\ \citenamefont
  {Kane}(2010)}]{HasanReview}%
  \BibitemOpen
  \bibfield  {author} {\bibinfo {author} {\bibfnamefont {M.~Z.}\ \bibnamefont
  {Hasan}}\ and\ \bibinfo {author} {\bibfnamefont {C.~L.}\ \bibnamefont
  {Kane}},\ }\href {\doibase 10.1103/RevModPhys.82.3045} {\bibfield  {journal}
  {\bibinfo  {journal} {Rev. Mod. Phys.}\ }\textbf {\bibinfo {volume} {82}},\
  \bibinfo {pages} {3045} (\bibinfo {year} {2010})}\BibitemShut {NoStop}%
\bibitem [{\citenamefont {Teo}\ and\ \citenamefont {Kane}(2010)}]{TeoKane}%
  \BibitemOpen
  \bibfield  {author} {\bibinfo {author} {\bibfnamefont {J.~C.~Y.}\
  \bibnamefont {Teo}}\ and\ \bibinfo {author} {\bibfnamefont {C.~L.}\
  \bibnamefont {Kane}},\ }\href {\doibase 10.1103/PhysRevB.82.115120}
  {\bibfield  {journal} {\bibinfo  {journal} {Phys. Rev. B}\ }\textbf {\bibinfo
  {volume} {82}},\ \bibinfo {pages} {115120} (\bibinfo {year}
  {2010})}\BibitemShut {NoStop}%
\bibitem [{\citenamefont {Thakurathi}\ \emph {et~al.}(2013)\citenamefont
  {Thakurathi}, \citenamefont {Patel}, \citenamefont {Sen},\ and\ \citenamefont
  {Dutta}}]{Thakurathi}%
  \BibitemOpen
  \bibfield  {author} {\bibinfo {author} {\bibfnamefont {M.}~\bibnamefont
  {Thakurathi}}, \bibinfo {author} {\bibfnamefont {A.~A.}\ \bibnamefont
  {Patel}}, \bibinfo {author} {\bibfnamefont {D.}~\bibnamefont {Sen}}, \ and\
  \bibinfo {author} {\bibfnamefont {A.}~\bibnamefont {Dutta}},\ }\href
  {\doibase 10.1103/PhysRevB.88.155133} {\bibfield  {journal} {\bibinfo
  {journal} {Phys. Rev. B}\ }\textbf {\bibinfo {volume} {88}},\ \bibinfo
  {pages} {155133} (\bibinfo {year} {2013})}\BibitemShut {NoStop}%
\bibitem [{\citenamefont {Rudner}\ \emph {et~al.}(2013)\citenamefont {Rudner},
  \citenamefont {Lindner}, \citenamefont {Berg},\ and\ \citenamefont
  {Levin}}]{Rudner13}%
  \BibitemOpen
  \bibfield  {author} {\bibinfo {author} {\bibfnamefont {M.~S.}\ \bibnamefont
  {Rudner}}, \bibinfo {author} {\bibfnamefont {N.~H.}\ \bibnamefont {Lindner}},
  \bibinfo {author} {\bibfnamefont {E.}~\bibnamefont {Berg}}, \ and\ \bibinfo
  {author} {\bibfnamefont {M.}~\bibnamefont {Levin}},\ }\href {\doibase
  10.1103/PhysRevX.3.031005} {\bibfield  {journal} {\bibinfo  {journal} {Phys.
  Rev. X}\ }\textbf {\bibinfo {volume} {3}},\ \bibinfo {pages} {031005}
  (\bibinfo {year} {2013})}\BibitemShut {NoStop}%
\bibitem [{\citenamefont {Khemani}\ \emph {et~al.}(2016)\citenamefont
  {Khemani}, \citenamefont {Lazarides}, \citenamefont {Moessner},\ and\
  \citenamefont {Sondhi}}]{Khemani}%
  \BibitemOpen
  \bibfield  {author} {\bibinfo {author} {\bibfnamefont {V.}~\bibnamefont
  {Khemani}}, \bibinfo {author} {\bibfnamefont {A.}~\bibnamefont {Lazarides}},
  \bibinfo {author} {\bibfnamefont {R.}~\bibnamefont {Moessner}}, \ and\
  \bibinfo {author} {\bibfnamefont {S.~L.}\ \bibnamefont {Sondhi}},\ }\href
  {\doibase 10.1103/PhysRevLett.116.250401} {\bibfield  {journal} {\bibinfo
  {journal} {Phys. Rev. Lett.}\ }\textbf {\bibinfo {volume} {116}},\ \bibinfo
  {pages} {250401} (\bibinfo {year} {2016})}\BibitemShut {NoStop}%
\bibitem [{\citenamefont {Else}\ \emph {et~al.}(2016)\citenamefont {Else},
  \citenamefont {Bauer},\ and\ \citenamefont {Nayak}}]{ElseTC}%
  \BibitemOpen
  \bibfield  {author} {\bibinfo {author} {\bibfnamefont {D.~V.}\ \bibnamefont
  {Else}}, \bibinfo {author} {\bibfnamefont {B.}~\bibnamefont {Bauer}}, \ and\
  \bibinfo {author} {\bibfnamefont {C.}~\bibnamefont {Nayak}},\ }\href
  {\doibase 10.1103/PhysRevLett.117.090402} {\bibfield  {journal} {\bibinfo
  {journal} {Phys. Rev. Lett.}\ }\textbf {\bibinfo {volume} {117}},\ \bibinfo
  {pages} {090402} (\bibinfo {year} {2016})}\BibitemShut {NoStop}%
\bibitem [{\citenamefont {Else}\ and\ \citenamefont {Nayak}(2016)}]{ElseFSPT}%
  \BibitemOpen
  \bibfield  {author} {\bibinfo {author} {\bibfnamefont {D.~V.}\ \bibnamefont
  {Else}}\ and\ \bibinfo {author} {\bibfnamefont {C.}~\bibnamefont {Nayak}},\
  }\href {\doibase 10.1103/PhysRevB.93.201103} {\bibfield  {journal} {\bibinfo
  {journal} {Phys. Rev. B}\ }\textbf {\bibinfo {volume} {93}},\ \bibinfo
  {pages} {201103} (\bibinfo {year} {2016})}\BibitemShut {NoStop}%
\bibitem [{\citenamefont {von Keyserlingk}\ and\ \citenamefont
  {Sondhi}(2016)}]{vonKeyserlingk}%
  \BibitemOpen
  \bibfield  {author} {\bibinfo {author} {\bibfnamefont {C.~W.}\ \bibnamefont
  {von Keyserlingk}}\ and\ \bibinfo {author} {\bibfnamefont {S.~L.}\
  \bibnamefont {Sondhi}},\ }\href {\doibase 10.1103/PhysRevB.93.245145}
  {\bibfield  {journal} {\bibinfo  {journal} {Phys. Rev. B}\ }\textbf {\bibinfo
  {volume} {93}},\ \bibinfo {pages} {245145} (\bibinfo {year}
  {2016})}\BibitemShut {NoStop}%
\bibitem [{\citenamefont {Nielsen}\ and\ \citenamefont
  {Ninomiya}(1981{\natexlab{a}})}]{Nielsen:1980rz}%
  \BibitemOpen
  \bibfield  {author} {\bibinfo {author} {\bibfnamefont {H.~B.}\ \bibnamefont
  {Nielsen}}\ and\ \bibinfo {author} {\bibfnamefont {M.}~\bibnamefont
  {Ninomiya}},\ }\href {\doibase 10.1016/0550-3213(82)90011-6} {\bibfield
  {journal} {\bibinfo  {journal} {Nucl. Phys. B}\ }\textbf {\bibinfo {volume}
  {185}},\ \bibinfo {pages} {20} (\bibinfo {year} {1981}{\natexlab{a}})},\
  \bibinfo {note} {[Erratum: Nucl.Phys.B 195, 541 (1982)]}\BibitemShut
  {NoStop}%
\bibitem [{\citenamefont {Nielsen}\ and\ \citenamefont
  {Ninomiya}(1981{\natexlab{b}})}]{Nielsen:1981xu}%
  \BibitemOpen
  \bibfield  {author} {\bibinfo {author} {\bibfnamefont {H.~B.}\ \bibnamefont
  {Nielsen}}\ and\ \bibinfo {author} {\bibfnamefont {M.}~\bibnamefont
  {Ninomiya}},\ }\href {\doibase 10.1016/0550-3213(81)90524-1} {\bibfield
  {journal} {\bibinfo  {journal} {Nucl. Phys. B}\ }\textbf {\bibinfo {volume}
  {193}},\ \bibinfo {pages} {173} (\bibinfo {year}
  {1981}{\natexlab{b}})}\BibitemShut {NoStop}%
\bibitem [{\citenamefont {B\"odeker}\ \emph {et~al.}(2000)\citenamefont
  {B\"odeker}, \citenamefont {Moore},\ and\ \citenamefont
  {Rummukainen}}]{GuyMoore}%
  \BibitemOpen
  \bibfield  {author} {\bibinfo {author} {\bibfnamefont {D.}~\bibnamefont
  {B\"odeker}}, \bibinfo {author} {\bibfnamefont {G.~D.}\ \bibnamefont
  {Moore}}, \ and\ \bibinfo {author} {\bibfnamefont {K.}~\bibnamefont
  {Rummukainen}},\ }\href {\doibase 10.1103/PhysRevD.61.056003} {\bibfield
  {journal} {\bibinfo  {journal} {Phys. Rev. D}\ }\textbf {\bibinfo {volume}
  {61}},\ \bibinfo {pages} {056003} (\bibinfo {year} {2000})}\BibitemShut
  {NoStop}%
\bibitem [{\citenamefont {Ambjorn}\ \emph {et~al.}(1991)\citenamefont
  {Ambjorn}, \citenamefont {Askgaard}, \citenamefont {Porter},\ and\
  \citenamefont {Shaposhnikov}}]{Ambjorn:1990pu}%
  \BibitemOpen
  \bibfield  {author} {\bibinfo {author} {\bibfnamefont {J.}~\bibnamefont
  {Ambjorn}}, \bibinfo {author} {\bibfnamefont {T.}~\bibnamefont {Askgaard}},
  \bibinfo {author} {\bibfnamefont {H.}~\bibnamefont {Porter}}, \ and\ \bibinfo
  {author} {\bibfnamefont {M.~E.}\ \bibnamefont {Shaposhnikov}},\ }\href
  {\doibase 10.1016/0550-3213(91)90341-T} {\bibfield  {journal} {\bibinfo
  {journal} {Nucl. Phys. B}\ }\textbf {\bibinfo {volume} {353}},\ \bibinfo
  {pages} {346} (\bibinfo {year} {1991})}\BibitemShut {NoStop}%
\bibitem [{\citenamefont {Aarts}\ and\ \citenamefont
  {Smit}(1999)}]{Aarts:1998td}%
  \BibitemOpen
  \bibfield  {author} {\bibinfo {author} {\bibfnamefont {G.}~\bibnamefont
  {Aarts}}\ and\ \bibinfo {author} {\bibfnamefont {J.}~\bibnamefont {Smit}},\
  }\href {\doibase 10.1016/S0550-3213(99)00320-X} {\bibfield  {journal}
  {\bibinfo  {journal} {Nucl. Phys. B}\ }\textbf {\bibinfo {volume} {555}},\
  \bibinfo {pages} {355} (\bibinfo {year} {1999})},\ \Eprint
  {http://arxiv.org/abs/hep-ph/9812413} {arXiv:hep-ph/9812413} \BibitemShut
  {NoStop}%
\bibitem [{\citenamefont {Mou}\ \emph {et~al.}(2013)\citenamefont {Mou},
  \citenamefont {Saffin},\ and\ \citenamefont {Tranberg}}]{Mou:2013kca}%
  \BibitemOpen
  \bibfield  {author} {\bibinfo {author} {\bibfnamefont {Z.-G.}\ \bibnamefont
  {Mou}}, \bibinfo {author} {\bibfnamefont {P.~M.}\ \bibnamefont {Saffin}}, \
  and\ \bibinfo {author} {\bibfnamefont {A.}~\bibnamefont {Tranberg}},\ }\href
  {\doibase 10.1007/JHEP11(2013)097} {\bibfield  {journal} {\bibinfo  {journal}
  {JHEP}\ }\textbf {\bibinfo {volume} {11}},\ \bibinfo {pages} {097} (\bibinfo
  {year} {2013})},\ \Eprint {http://arxiv.org/abs/1307.7924} {arXiv:1307.7924
  [hep-ph]} \BibitemShut {NoStop}%
\bibitem [{\citenamefont {Iadecola}\ \emph {et~al.}(2023)\citenamefont
  {Iadecola}, \citenamefont {Sen},\ and\ \citenamefont
  {Sivertsen}}]{FloquetLattice1}%
  \BibitemOpen
  \bibfield  {author} {\bibinfo {author} {\bibfnamefont {T.}~\bibnamefont
  {Iadecola}}, \bibinfo {author} {\bibfnamefont {S.}~\bibnamefont {Sen}}, \
  and\ \bibinfo {author} {\bibfnamefont {L.}~\bibnamefont {Sivertsen}},\ }\href
  {\doibase 10.48550/ARXIV.2306.16463} {\enquote {\bibinfo {title} {Floquet
  insulators and lattice fermions},}\ } (\bibinfo {year} {2023})\BibitemShut
  {NoStop}%
\bibitem [{\citenamefont {Su}\ \emph {et~al.}(1979)\citenamefont {Su},
  \citenamefont {Schrieffer},\ and\ \citenamefont {Heeger}}]{SSH}%
  \BibitemOpen
  \bibfield  {author} {\bibinfo {author} {\bibfnamefont {W.~P.}\ \bibnamefont
  {Su}}, \bibinfo {author} {\bibfnamefont {J.~R.}\ \bibnamefont {Schrieffer}},
  \ and\ \bibinfo {author} {\bibfnamefont {A.~J.}\ \bibnamefont {Heeger}},\
  }\href {\doibase 10.1103/PhysRevLett.42.1698} {\bibfield  {journal} {\bibinfo
   {journal} {Phys. Rev. Lett.}\ }\textbf {\bibinfo {volume} {42}},\ \bibinfo
  {pages} {1698} (\bibinfo {year} {1979})}\BibitemShut {NoStop}%
\bibitem [{\citenamefont {Kaplan}(1992)}]{Kaplan:1992bt}%
  \BibitemOpen
  \bibfield  {author} {\bibinfo {author} {\bibfnamefont {D.~B.}\ \bibnamefont
  {Kaplan}},\ }\href {\doibase 10.1016/0370-2693(92)91112-M} {\bibfield
  {journal} {\bibinfo  {journal} {Phys. Lett. B}\ }\textbf {\bibinfo {volume}
  {288}},\ \bibinfo {pages} {342} (\bibinfo {year} {1992})},\ \Eprint
  {http://arxiv.org/abs/hep-lat/9206013} {arXiv:hep-lat/9206013} \BibitemShut
  {NoStop}%
\bibitem [{\citenamefont {Shamir}(1993)}]{Shamir:1993zy}%
  \BibitemOpen
  \bibfield  {author} {\bibinfo {author} {\bibfnamefont {Y.}~\bibnamefont
  {Shamir}},\ }\href {\doibase 10.1016/0550-3213(93)90162-I} {\bibfield
  {journal} {\bibinfo  {journal} {Nucl. Phys. B}\ }\textbf {\bibinfo {volume}
  {406}},\ \bibinfo {pages} {90} (\bibinfo {year} {1993})},\ \Eprint
  {http://arxiv.org/abs/hep-lat/9303005} {arXiv:hep-lat/9303005} \BibitemShut
  {NoStop}%
\bibitem [{\citenamefont {Ginsparg}\ and\ \citenamefont
  {Wilson}(1982)}]{PhysRevD.25.2649}%
  \BibitemOpen
  \bibfield  {author} {\bibinfo {author} {\bibfnamefont {P.~H.}\ \bibnamefont
  {Ginsparg}}\ and\ \bibinfo {author} {\bibfnamefont {K.~G.}\ \bibnamefont
  {Wilson}},\ }\href {\doibase 10.1103/PhysRevD.25.2649} {\bibfield  {journal}
  {\bibinfo  {journal} {Phys. Rev. D}\ }\textbf {\bibinfo {volume} {25}},\
  \bibinfo {pages} {2649} (\bibinfo {year} {1982})}\BibitemShut {NoStop}%
\bibitem [{\citenamefont {Neuberger}(1998{\natexlab{a}})}]{Neuberger:1997fp}%
  \BibitemOpen
  \bibfield  {author} {\bibinfo {author} {\bibfnamefont {H.}~\bibnamefont
  {Neuberger}},\ }\href {\doibase 10.1016/S0370-2693(97)01368-3} {\bibfield
  {journal} {\bibinfo  {journal} {Phys. Lett. B}\ }\textbf {\bibinfo {volume}
  {417}},\ \bibinfo {pages} {141} (\bibinfo {year} {1998}{\natexlab{a}})},\
  \Eprint {http://arxiv.org/abs/hep-lat/9707022} {arXiv:hep-lat/9707022}
  \BibitemShut {NoStop}%
\bibitem [{\citenamefont {Neuberger}(1998{\natexlab{b}})}]{Neuberger:1998wv}%
  \BibitemOpen
  \bibfield  {author} {\bibinfo {author} {\bibfnamefont {H.}~\bibnamefont
  {Neuberger}},\ }\href {\doibase 10.1016/S0370-2693(98)00355-4} {\bibfield
  {journal} {\bibinfo  {journal} {Phys. Lett. B}\ }\textbf {\bibinfo {volume}
  {427}},\ \bibinfo {pages} {353} (\bibinfo {year} {1998}{\natexlab{b}})},\
  \Eprint {http://arxiv.org/abs/hep-lat/9801031} {arXiv:hep-lat/9801031}
  \BibitemShut {NoStop}%
\bibitem [{\citenamefont {Kogut}\ and\ \citenamefont
  {Susskind}(1975)}]{PhysRevD.11.395}%
  \BibitemOpen
  \bibfield  {author} {\bibinfo {author} {\bibfnamefont {J.}~\bibnamefont
  {Kogut}}\ and\ \bibinfo {author} {\bibfnamefont {L.}~\bibnamefont
  {Susskind}},\ }\href {\doibase 10.1103/PhysRevD.11.395} {\bibfield  {journal}
  {\bibinfo  {journal} {Phys. Rev. D}\ }\textbf {\bibinfo {volume} {11}},\
  \bibinfo {pages} {395} (\bibinfo {year} {1975})}\BibitemShut {NoStop}%
\bibitem [{\citenamefont {Jackiw}\ and\ \citenamefont
  {Rebbi}(1976)}]{JackiwRebbi}%
  \BibitemOpen
  \bibfield  {author} {\bibinfo {author} {\bibfnamefont {R.}~\bibnamefont
  {Jackiw}}\ and\ \bibinfo {author} {\bibfnamefont {C.}~\bibnamefont {Rebbi}},\
  }\href {\doibase 10.1103/PhysRevD.13.3398} {\bibfield  {journal} {\bibinfo
  {journal} {Phys. Rev. D}\ }\textbf {\bibinfo {volume} {13}},\ \bibinfo
  {pages} {3398} (\bibinfo {year} {1976})}\BibitemShut {NoStop}%
\bibitem [{\citenamefont {Potter}\ \emph {et~al.}(2016)\citenamefont {Potter},
  \citenamefont {Morimoto},\ and\ \citenamefont {Vishwanath}}]{Potter}%
  \BibitemOpen
  \bibfield  {author} {\bibinfo {author} {\bibfnamefont {A.~C.}\ \bibnamefont
  {Potter}}, \bibinfo {author} {\bibfnamefont {T.}~\bibnamefont {Morimoto}}, \
  and\ \bibinfo {author} {\bibfnamefont {A.}~\bibnamefont {Vishwanath}},\
  }\href {\doibase 10.1103/PhysRevX.6.041001} {\bibfield  {journal} {\bibinfo
  {journal} {Phys. Rev. X}\ }\textbf {\bibinfo {volume} {6}},\ \bibinfo {pages}
  {041001} (\bibinfo {year} {2016})}\BibitemShut {NoStop}%
\bibitem [{Note1()}]{Note1}%
  \BibitemOpen
  \bibinfo {note} {In theories with Euclidean rotational invariance, $R_{\mu }$
  is $\mu $ independent and set to $R_{\mu }=R$.}\BibitemShut {Stop}%
\bibitem [{Note2()}]{Note2}%
  \BibitemOpen
  \bibinfo {note} {Note that, in previous work, we had mapped the Floquet
  spectrum to Minkowski space lattice Fermions using the same condition. In
  that case, the zeroes of the Minkowski Fermion operator match one to one with
  the zeroes of the Floquet Schr{\"o}dinger operator $i\partial _t-H_F$. The
  solutions to the equation \protect \eqref {eq:ff't} in Euclidean space are
  not necessarily zeroes of the fermion operator unless $i\protect \mathcal
  {F}=\protect \mathcal {F}'=0$. This is the only distinction between the
  Euclidean and Minkowski constructions and it does not affect any of our
  analysis.}\BibitemShut {Stop}%
\bibitem [{\citenamefont {D'Alessio}\ and\ \citenamefont
  {Rigol}(2014)}]{D'Alessio}%
  \BibitemOpen
  \bibfield  {author} {\bibinfo {author} {\bibfnamefont {L.}~\bibnamefont
  {D'Alessio}}\ and\ \bibinfo {author} {\bibfnamefont {M.}~\bibnamefont
  {Rigol}},\ }\href {\doibase 10.1103/PhysRevX.4.041048} {\bibfield  {journal}
  {\bibinfo  {journal} {Phys. Rev. X}\ }\textbf {\bibinfo {volume} {4}},\
  \bibinfo {pages} {041048} (\bibinfo {year} {2014})}\BibitemShut {NoStop}%
\bibitem [{\citenamefont {Abanin}\ \emph {et~al.}(2017)\citenamefont {Abanin},
  \citenamefont {Roeck}, \citenamefont {Ho},\ and\ \citenamefont
  {Huveneers}}]{Abanin17}%
  \BibitemOpen
  \bibfield  {author} {\bibinfo {author} {\bibfnamefont {D.}~\bibnamefont
  {Abanin}}, \bibinfo {author} {\bibfnamefont {W.~D.}\ \bibnamefont {Roeck}},
  \bibinfo {author} {\bibfnamefont {W.~W.}\ \bibnamefont {Ho}}, \ and\ \bibinfo
  {author} {\bibfnamefont {F.}~\bibnamefont {Huveneers}},\ }\href {\doibase
  10.1007/s00220-017-2930-x} {\bibfield  {journal} {\bibinfo  {journal}
  {Communications in Mathematical Physics}\ }\textbf {\bibinfo {volume}
  {354}},\ \bibinfo {pages} {809} (\bibinfo {year} {2017})}\BibitemShut
  {NoStop}%
\bibitem [{\citenamefont {Kitagawa}\ \emph {et~al.}(2010)\citenamefont
  {Kitagawa}, \citenamefont {Berg}, \citenamefont {Rudner},\ and\ \citenamefont
  {Demler}}]{Kitagawa}%
  \BibitemOpen
  \bibfield  {author} {\bibinfo {author} {\bibfnamefont {T.}~\bibnamefont
  {Kitagawa}}, \bibinfo {author} {\bibfnamefont {E.}~\bibnamefont {Berg}},
  \bibinfo {author} {\bibfnamefont {M.}~\bibnamefont {Rudner}}, \ and\ \bibinfo
  {author} {\bibfnamefont {E.}~\bibnamefont {Demler}},\ }\href {\doibase
  10.1103/PhysRevB.82.235114} {\bibfield  {journal} {\bibinfo  {journal} {Phys.
  Rev. B}\ }\textbf {\bibinfo {volume} {82}},\ \bibinfo {pages} {235114}
  (\bibinfo {year} {2010})}\BibitemShut {NoStop}%
\bibitem [{\citenamefont {Nathan}\ and\ \citenamefont {Rudner}(2015)}]{Nathan}%
  \BibitemOpen
  \bibfield  {author} {\bibinfo {author} {\bibfnamefont {F.}~\bibnamefont
  {Nathan}}\ and\ \bibinfo {author} {\bibfnamefont {M.~S.}\ \bibnamefont
  {Rudner}},\ }\href {\doibase 10.1088/1367-2630/17/12/125014} {\bibfield
  {journal} {\bibinfo  {journal} {New Journal of Physics}\ }\textbf {\bibinfo
  {volume} {17}},\ \bibinfo {pages} {125014} (\bibinfo {year}
  {2015})}\BibitemShut {NoStop}%
\bibitem [{\citenamefont {Carpentier}\ \emph {et~al.}(2015)\citenamefont
  {Carpentier}, \citenamefont {Delplace}, \citenamefont {Fruchart},\ and\
  \citenamefont {Gaw\ifmmode{\c{e}}\else{\c{e}}\fi{}dzki}}]{Carpentier}%
  \BibitemOpen
  \bibfield  {author} {\bibinfo {author} {\bibfnamefont {D.}~\bibnamefont
  {Carpentier}}, \bibinfo {author} {\bibfnamefont {P.}~\bibnamefont
  {Delplace}}, \bibinfo {author} {\bibfnamefont {M.}~\bibnamefont {Fruchart}},
  \ and\ \bibinfo {author} {\bibfnamefont {K.}~\bibnamefont
  {Gaw\ifmmode{\c{e}}\else{\c{e}}\fi{}dzki}},\ }\href {\doibase
  10.1103/PhysRevLett.114.106806} {\bibfield  {journal} {\bibinfo  {journal}
  {Phys. Rev. Lett.}\ }\textbf {\bibinfo {volume} {114}},\ \bibinfo {pages}
  {106806} (\bibinfo {year} {2015})}\BibitemShut {NoStop}%
\bibitem [{\citenamefont {Fruchart}(2016)}]{Fruchart}%
  \BibitemOpen
  \bibfield  {author} {\bibinfo {author} {\bibfnamefont {M.}~\bibnamefont
  {Fruchart}},\ }\href {\doibase 10.1103/PhysRevB.93.115429} {\bibfield
  {journal} {\bibinfo  {journal} {Phys. Rev. B}\ }\textbf {\bibinfo {volume}
  {93}},\ \bibinfo {pages} {115429} (\bibinfo {year} {2016})}\BibitemShut
  {NoStop}%
\bibitem [{\citenamefont {Golterman}\ \emph {et~al.}(1993)\citenamefont
  {Golterman}, \citenamefont {Jansen},\ and\ \citenamefont
  {Kaplan}}]{Golterman:1992ub}%
  \BibitemOpen
  \bibfield  {author} {\bibinfo {author} {\bibfnamefont {M.~F.~L.}\
  \bibnamefont {Golterman}}, \bibinfo {author} {\bibfnamefont {K.}~\bibnamefont
  {Jansen}}, \ and\ \bibinfo {author} {\bibfnamefont {D.~B.}\ \bibnamefont
  {Kaplan}},\ }\href {\doibase 10.1016/0370-2693(93)90692-B} {\bibfield
  {journal} {\bibinfo  {journal} {Phys. Lett. B}\ }\textbf {\bibinfo {volume}
  {301}},\ \bibinfo {pages} {219} (\bibinfo {year} {1993})},\ \Eprint
  {http://arxiv.org/abs/hep-lat/9209003} {arXiv:hep-lat/9209003} \BibitemShut
  {NoStop}%
\bibitem [{\citenamefont {Bermudez}\ \emph {et~al.}(2017)\citenamefont
  {Bermudez}, \citenamefont {Tagliacozzo}, \citenamefont {Sierra},\ and\
  \citenamefont {Richerme}}]{PhysRevB.95.024431}%
  \BibitemOpen
  \bibfield  {author} {\bibinfo {author} {\bibfnamefont {A.}~\bibnamefont
  {Bermudez}}, \bibinfo {author} {\bibfnamefont {L.}~\bibnamefont
  {Tagliacozzo}}, \bibinfo {author} {\bibfnamefont {G.}~\bibnamefont {Sierra}},
  \ and\ \bibinfo {author} {\bibfnamefont {P.}~\bibnamefont {Richerme}},\
  }\href {\doibase 10.1103/PhysRevB.95.024431} {\bibfield  {journal} {\bibinfo
  {journal} {Phys. Rev. B}\ }\textbf {\bibinfo {volume} {95}},\ \bibinfo
  {pages} {024431} (\bibinfo {year} {2017})}\BibitemShut {NoStop}%
\bibitem [{\citenamefont {Scholl}\ \emph {et~al.}(2022)\citenamefont {Scholl},
  \citenamefont {Williams}, \citenamefont {Bornet}, \citenamefont {Wallner},
  \citenamefont {Barredo}, \citenamefont {Henriet}, \citenamefont {Signoles},
  \citenamefont {Hainaut}, \citenamefont {Franz}, \citenamefont {Geier},
  \citenamefont {Tebben}, \citenamefont {Salzinger}, \citenamefont {Z\"urn},
  \citenamefont {Lahaye}, \citenamefont {Weidem\"uller},\ and\ \citenamefont
  {Browaeys}}]{PRXQuantum.3.020303}%
  \BibitemOpen
  \bibfield  {author} {\bibinfo {author} {\bibfnamefont {P.}~\bibnamefont
  {Scholl}}, \bibinfo {author} {\bibfnamefont {H.~J.}\ \bibnamefont
  {Williams}}, \bibinfo {author} {\bibfnamefont {G.}~\bibnamefont {Bornet}},
  \bibinfo {author} {\bibfnamefont {F.}~\bibnamefont {Wallner}}, \bibinfo
  {author} {\bibfnamefont {D.}~\bibnamefont {Barredo}}, \bibinfo {author}
  {\bibfnamefont {L.}~\bibnamefont {Henriet}}, \bibinfo {author} {\bibfnamefont
  {A.}~\bibnamefont {Signoles}}, \bibinfo {author} {\bibfnamefont
  {C.}~\bibnamefont {Hainaut}}, \bibinfo {author} {\bibfnamefont
  {T.}~\bibnamefont {Franz}}, \bibinfo {author} {\bibfnamefont
  {S.}~\bibnamefont {Geier}}, \bibinfo {author} {\bibfnamefont
  {A.}~\bibnamefont {Tebben}}, \bibinfo {author} {\bibfnamefont
  {A.}~\bibnamefont {Salzinger}}, \bibinfo {author} {\bibfnamefont
  {G.}~\bibnamefont {Z\"urn}}, \bibinfo {author} {\bibfnamefont
  {T.}~\bibnamefont {Lahaye}}, \bibinfo {author} {\bibfnamefont
  {M.}~\bibnamefont {Weidem\"uller}}, \ and\ \bibinfo {author} {\bibfnamefont
  {A.}~\bibnamefont {Browaeys}},\ }\href {\doibase 10.1103/PRXQuantum.3.020303}
  {\bibfield  {journal} {\bibinfo  {journal} {PRX Quantum}\ }\textbf {\bibinfo
  {volume} {3}},\ \bibinfo {pages} {020303} (\bibinfo {year}
  {2022})}\BibitemShut {NoStop}%
\bibitem [{\citenamefont {Choi}\ \emph {et~al.}(2020)\citenamefont {Choi},
  \citenamefont {Zhou}, \citenamefont {Knowles}, \citenamefont {Landig},
  \citenamefont {Choi},\ and\ \citenamefont {Lukin}}]{PhysRevX.10.031002}%
  \BibitemOpen
  \bibfield  {author} {\bibinfo {author} {\bibfnamefont {J.}~\bibnamefont
  {Choi}}, \bibinfo {author} {\bibfnamefont {H.}~\bibnamefont {Zhou}}, \bibinfo
  {author} {\bibfnamefont {H.~S.}\ \bibnamefont {Knowles}}, \bibinfo {author}
  {\bibfnamefont {R.}~\bibnamefont {Landig}}, \bibinfo {author} {\bibfnamefont
  {S.}~\bibnamefont {Choi}}, \ and\ \bibinfo {author} {\bibfnamefont {M.~D.}\
  \bibnamefont {Lukin}},\ }\href {\doibase 10.1103/PhysRevX.10.031002}
  {\bibfield  {journal} {\bibinfo  {journal} {Phys. Rev. X}\ }\textbf {\bibinfo
  {volume} {10}},\ \bibinfo {pages} {031002} (\bibinfo {year}
  {2020})}\BibitemShut {NoStop}%
\bibitem [{\citenamefont {Wintersperger}\ \emph {et~al.}(2020)\citenamefont
  {Wintersperger}, \citenamefont {Braun}, \citenamefont {Ünal}, \citenamefont
  {Eckardt}, \citenamefont {Liberto}, \citenamefont {Goldman}, \citenamefont
  {Bloch},\ and\ \citenamefont {Aidelsburger}}]{Wintersperger_2020}%
  \BibitemOpen
  \bibfield  {author} {\bibinfo {author} {\bibfnamefont {K.}~\bibnamefont
  {Wintersperger}}, \bibinfo {author} {\bibfnamefont {C.}~\bibnamefont
  {Braun}}, \bibinfo {author} {\bibfnamefont {F.~N.}\ \bibnamefont {Ünal}},
  \bibinfo {author} {\bibfnamefont {A.}~\bibnamefont {Eckardt}}, \bibinfo
  {author} {\bibfnamefont {M.~D.}\ \bibnamefont {Liberto}}, \bibinfo {author}
  {\bibfnamefont {N.}~\bibnamefont {Goldman}}, \bibinfo {author} {\bibfnamefont
  {I.}~\bibnamefont {Bloch}}, \ and\ \bibinfo {author} {\bibfnamefont
  {M.}~\bibnamefont {Aidelsburger}},\ }\href {\doibase
  10.1038/s41567-020-0949-y} {\bibfield  {journal} {\bibinfo  {journal} {Nature
  Physics}\ }\textbf {\bibinfo {volume} {16}},\ \bibinfo {pages} {1058}
  (\bibinfo {year} {2020})}\BibitemShut {NoStop}%
\bibitem [{\citenamefont {Ciavarella}\ \emph {et~al.}(2023)\citenamefont
  {Ciavarella}, \citenamefont {Caspar}, \citenamefont {Singh}, \citenamefont
  {Savage},\ and\ \citenamefont {Lougovski}}]{Ciavarella:2022tvc}%
  \BibitemOpen
  \bibfield  {author} {\bibinfo {author} {\bibfnamefont {A.~N.}\ \bibnamefont
  {Ciavarella}}, \bibinfo {author} {\bibfnamefont {S.}~\bibnamefont {Caspar}},
  \bibinfo {author} {\bibfnamefont {H.}~\bibnamefont {Singh}}, \bibinfo
  {author} {\bibfnamefont {M.~J.}\ \bibnamefont {Savage}}, \ and\ \bibinfo
  {author} {\bibfnamefont {P.}~\bibnamefont {Lougovski}},\ }\href {\doibase
  10.1103/PhysRevA.108.042216} {\bibfield  {journal} {\bibinfo  {journal}
  {Phys. Rev. A}\ }\textbf {\bibinfo {volume} {108}},\ \bibinfo {pages}
  {042216} (\bibinfo {year} {2023})},\ \Eprint
  {http://arxiv.org/abs/2207.09438} {arXiv:2207.09438 [quant-ph]} \BibitemShut
  {NoStop}%
\bibitem [{\citenamefont {Schweizer}\ \emph {et~al.}(2019)\citenamefont
  {Schweizer}, \citenamefont {Grusdt}, \citenamefont {Berngruber},
  \citenamefont {Barbiero}, \citenamefont {Demler}, \citenamefont {Goldman},
  \citenamefont {Bloch},\ and\ \citenamefont {Aidelsburger}}]{Schweizer_2019}%
  \BibitemOpen
  \bibfield  {author} {\bibinfo {author} {\bibfnamefont {C.}~\bibnamefont
  {Schweizer}}, \bibinfo {author} {\bibfnamefont {F.}~\bibnamefont {Grusdt}},
  \bibinfo {author} {\bibfnamefont {M.}~\bibnamefont {Berngruber}}, \bibinfo
  {author} {\bibfnamefont {L.}~\bibnamefont {Barbiero}}, \bibinfo {author}
  {\bibfnamefont {E.}~\bibnamefont {Demler}}, \bibinfo {author} {\bibfnamefont
  {N.}~\bibnamefont {Goldman}}, \bibinfo {author} {\bibfnamefont
  {I.}~\bibnamefont {Bloch}}, \ and\ \bibinfo {author} {\bibfnamefont
  {M.}~\bibnamefont {Aidelsburger}},\ }\href {\doibase
  10.1038/s41567-019-0649-7} {\bibfield  {journal} {\bibinfo  {journal} {Nature
  Physics}\ }\textbf {\bibinfo {volume} {15}},\ \bibinfo {pages} {1168}
  (\bibinfo {year} {2019})}\BibitemShut {NoStop}%
\bibitem [{\citenamefont {Goldman}\ and\ \citenamefont
  {Dalibard}(2014)}]{PhysRevX.4.031027}%
  \BibitemOpen
  \bibfield  {author} {\bibinfo {author} {\bibfnamefont {N.}~\bibnamefont
  {Goldman}}\ and\ \bibinfo {author} {\bibfnamefont {J.}~\bibnamefont
  {Dalibard}},\ }\href {\doibase 10.1103/PhysRevX.4.031027} {\bibfield
  {journal} {\bibinfo  {journal} {Phys. Rev. X}\ }\textbf {\bibinfo {volume}
  {4}},\ \bibinfo {pages} {031027} (\bibinfo {year} {2014})}\BibitemShut
  {NoStop}%
\bibitem [{\citenamefont {Aidelsburger}\ \emph {et~al.}(2013)\citenamefont
  {Aidelsburger}, \citenamefont {Atala}, \citenamefont {Lohse}, \citenamefont
  {Barreiro}, \citenamefont {Paredes},\ and\ \citenamefont
  {Bloch}}]{PhysRevLett.111.185301}%
  \BibitemOpen
  \bibfield  {author} {\bibinfo {author} {\bibfnamefont {M.}~\bibnamefont
  {Aidelsburger}}, \bibinfo {author} {\bibfnamefont {M.}~\bibnamefont {Atala}},
  \bibinfo {author} {\bibfnamefont {M.}~\bibnamefont {Lohse}}, \bibinfo
  {author} {\bibfnamefont {J.~T.}\ \bibnamefont {Barreiro}}, \bibinfo {author}
  {\bibfnamefont {B.}~\bibnamefont {Paredes}}, \ and\ \bibinfo {author}
  {\bibfnamefont {I.}~\bibnamefont {Bloch}},\ }\href {\doibase
  10.1103/PhysRevLett.111.185301} {\bibfield  {journal} {\bibinfo  {journal}
  {Phys. Rev. Lett.}\ }\textbf {\bibinfo {volume} {111}},\ \bibinfo {pages}
  {185301} (\bibinfo {year} {2013})}\BibitemShut {NoStop}%
\bibitem [{\citenamefont {Eckardt}(2017)}]{RevModPhys.89.011004}%
  \BibitemOpen
  \bibfield  {author} {\bibinfo {author} {\bibfnamefont {A.}~\bibnamefont
  {Eckardt}},\ }\href {\doibase 10.1103/RevModPhys.89.011004} {\bibfield
  {journal} {\bibinfo  {journal} {Rev. Mod. Phys.}\ }\textbf {\bibinfo {volume}
  {89}},\ \bibinfo {pages} {011004} (\bibinfo {year} {2017})}\BibitemShut
  {NoStop}%
\end{thebibliography}%

\end{document}